\documentclass[a4paper,11pt]{article}
\pdfoutput=1 

\usepackage{jheppub} 
\usepackage[export]{adjustbox}
\usepackage[utf8]{inputenc}
\usepackage{multirow}
\usepackage{bbold}
\usepackage[table]{xcolor}
\usepackage{slashed}
\usepackage{bm}
\usepackage{subfig}
\usepackage{float}
\usepackage{fancyvrb}
\usepackage{fvextra}
\usepackage{soul}
\usepackage[title]{appendix}
\usepackage{physics}
\usepackage{siunitx}
\usepackage{dsfont}
\usepackage{listings}
\usepackage{ulem}
\usepackage{xspace}

\definecolor{myg}{RGB}{56, 140, 70}

\usepackage{subfig}
\usepackage{color,graphicx,slashed,hyperref}
\usepackage[utf8]{inputenc}
\hypersetup{
   bookmarks=true,         
   unicode=true,          
   pdftoolbar=true,        
   pdfmenubar=true,        
   pdffitwindow=false,     
   pdfstartview={FitH},    
   pdftitle={LeptonFlavourTransfers},    
   pdfauthor={Author},     
   pdfsubject={Subject},   
   pdfcreator={Creator},   
   pdfproducer={Producer}, 
   pdfkeywords={keyword1} {key2} {key3}, 
   pdfnewwindow=true,      
   colorlinks=true,       
   linkcolor=black,          
   citecolor=black,        
   filecolor=black,      
   urlcolor=black           
}

\def\Wt{V_{p,m}}

\def\Zt{V_3}

\newcommand{\marty}{\texttt{MARTY}}
\newcommand{\darkpack}{\texttt{DarkPACK}\xspace}
\newcommand{\darksusy}{\texttt{DarkSUSY}\xspace}

\newcommand{\mtx}[1]{\begin{pmatrix}#1\end{pmatrix}}
\newcommand{\amc}{{\sc MadGraph5}\_a{\sc MC@NLO}}

\newcommand{\FVem}{$(\textrm{FV})_{12}$}
\newcommand{\FVet}{$(\textrm{FV})_{13}$}
\newcommand{\FVmt}{$(\textrm{FV})_{23}$}
\newcommand{\FTem}{$(\textrm{FT})_{12}$}
\newcommand{\FTet}{$(\textrm{FT})_{13}$}
\newcommand{\FTmt}{$(\textrm{FT})_{23}$}

\title{Leptonic flavour transfer: a new window on flavour gauge symmetries}

\author[a]{L. Darm\'e,}
\emailAdd{l.darme@ip2i.in2p3.fr}
\author[a]{N. Fardeau,}
\emailAdd{n.fardeau@ip2i.in2p3.fr}
\author[a,b,c]{F. Mahmoudi,}
\emailAdd{nazila@cern.ch}
\author[d]{M. Palmiotto}
\emailAdd{marco.palmiotto@fys.uio.no}

\affiliation[a]{\vspace*{0.5cm}Université Lyon 1, CNRS, IP2I, UMR 5822, Villeurbanne, France}
\affiliation[b]{Theoretical Physics Department, CERN, CH-1211 Geneva 23, Switzerland}
\affiliation[c]{Institut Universitaire de France (IUF), 75005 Paris, France}
\affiliation[d]{University of Oslo, Department of Physics, 0316 Oslo, Norway\vspace*{0.5cm} }

\abstract{
New flavour non-abelian gauge groups, which may arise as part of a fundamental theory of flavour, can lead to distinctive flavour-transfer processes. When restricted to the lepton sector, such processes partially mimic the standard charged current interactions at low energy. We explicitly study such constructions with various flavour structures, and investigate systematically all relevant accelerator-based constraints, exploring specific experimental signatures for these models. In the absence of flavour-breaking spurions, constraints from heavy lepton lifetime are found to dominate over most of the parameter space, with specific parts still open for future ee-collider searches. Numerical predictions are obtained using the \texttt{MARTY} framework, allowing us to consistently explore both the light- and heavy-mediator regimes. Finally, using \darkpack, we also assessed that extensions of such models could be compatible with dark matter relic density bounds. 
}


\begin{document}

\maketitle

\setcounter{footnote}{0}

\newenvironment{Appendix}
{
	\setcounter{section}{1}
	\setcounter{equation}{0}
	\renewcommand{\thesubsection}{\Alph{subsection}}
	\renewcommand{\theequation}{A.\arabic{equation}}
}

\section{Introduction\label{sec:intro}}

Flavour remains one of the least understood aspects of the Standard Model (SM). While the gauge interactions are highly constrained by symmetry principles, the observed pattern of fermion masses and mixings along with their hierarchical structure are not explained within the SM. This has motivated a broad class of extensions in which flavour is promoted from an accidental structure of the low-energy theory to a consequence of an underlying symmetry~\cite{Fritzsch:1999ee,Xing:2014sja,Feruglio:2015jfa}.

Among these possibilities, New Physics (NP) scenarios based on gauge extensions acting in flavour space are particularly appealing because they introduce additional structure rather than merely enlarging the particle spectrum~\cite{Monich:1980rr,Berezhiani:1983hm}.
However, anomaly cancellation and the observed suppression of flavour-changing processes strongly restrict the viable constructions. In the abelian case, anomaly-free realisations typically require non-universal flavour charges, while non-abelian flavour symmetries are even more constrained by the assignment of fermions into flavour multiplets.

One of the simplest non-abelian possibilities consists in introducing a horizontal gauge flavour symmetry acting on fermion generations~\cite{Monich:1980rr,Berezhiani:1990wn,Chiang:2017vcl,Pomarol:1995xc,Barbieri:1995uv,Barbieri:2011ci,Guadagnoli:2018ojc,Belfatto:2018cfo,Belfatto:2019swo,Carvunis:2020exc,Greljo:2023bix,Darme:2022uzl,Antusch:2023shi,Greljo:2024zrj,Calibbi:2025rxn}.
In such constructions, new gauge bosons mediate transitions between the fermions belonging to the same flavour multiplet. After rotating to the mass basis, these interactions generate a characteristic pattern of flavour observables whose structure differs from conventional flavour-changing neutral current scenarios.

In this work, we focus on the purely leptonic case and consider an additional horizontal $SU(2)_f$ gauge symmetry acting on the lepton sector. A distinctive feature of this setup is the appearance of {\it flavour-transfer} (FT) processes, introduced in \cite{Darme:2023nsy}, in which flavour is exchanged between leptonic states through the new gauge interaction while remaining compatible with the underlying flavour symmetry. In the limit where the charged-lepton mass basis is aligned with the flavour gauge basis, these processes survive whereas more conventional flavour-violating observables are parametrically suppressed~\cite{Carvunis:2020exc}.

Previous studies have explored the consequences of flavour gauge symmetries in several contexts, including flavour textures, low-energy flavour observables and simplified phenomenological descriptions of flavourful mediators. In earlier work~\cite{Darme:2023nsy}, the notion of flavour transfer was introduced as a characteristic consequence of non-abelian flavour interactions, highlighting how flavour exchange processes can remain observable even in regimes where conventional flavour-violating signatures are suppressed. The present work builds on this idea and focuses on its realisation in the leptonic sector.

This structure leads to a phenomenology that is distinct from conventional lepton flavour violation~\cite{Kuno:1999jp,Calibbi:2017uvl}. Since neutrinos naturally appear in multiple flavour states and remain experimentally indistinguishable in many final states, neutrino experiments~ \cite{Farzan:2017xzy,Ohlsson:2012kf}  provide sensitive probes of flavour transfer. At the same time, departures from exact flavour alignment induce additional signals in rare lepton decays, precision observables and radiative effects. 

The goal of this article is to study systematically the phenomenology of leptonic flavour transfer mediated by flavour gauge bosons with or without additional flavour-breaking small spurions. To this end, we have established a broad classification of the possible effective low-energy interactions arising from the underlying $SU(2)_f$ structure, exploring additionally a $U(2)_f$ structure leading to the proper leptonic mass spectrum and the corresponding low-energy interactions which leads to specific unsuppressed mixings and thus to more generic flavour-violating processes. We further briefly consider the possible connections with dark-sector phenomenology and astrophysical observations. 

The technical core of this work lies then in the systematic study of the constraints from laboratory experiments in the case where the new flavour gauge bosons are light enough that an effective theory approach breaks down. This significantly enriches the corresponding phenomenology due to the various mass thresholds. In order to treat numerically this problem, we build on the \marty~\cite{uhlrich__2021,Uhlrich:2020aaj,Uhlrich:2021ded} framework, extending its capability from estimated effective operators and squared amplitude towards obtaining decay widths or scattering cross-sections. This allows us to identify several specific predictions of low-mass flavour-transfer processes (such as flavour-mixed $e\mu$ visible leptons final states in neutrino trident processes) and to re-interpret existing constraints to our models.  We then consider the phenomenological consequences of adding flavour-violating spurions to the $SU(2)_f$ structure, similarly leveraging the newly developed capabilities of~\marty~ in order to generate all possible sets of rare flavour-violating $\mu$ and $\tau$ decays with possible emission of an on-shell flavour gauge boson. Expectedly, the corresponding constraints are extremely stringent and tend to restrict the gauge couplings to the $10^{-9}$ range at low masses. Finally, we study several relevant one-loop processes, including in particular the various lepton anomalous magnetic moments. For these last observables and despite a contrasted experimental situation, we find that the phenomenologically-relevant regions with a new physics contribution of $|a_\mu| \sim 10^{-9}$ and $|a_e| \sim 10^{-12}$ are in certain cases still compatible with other bounds. 
We further add a Dirac fermion doublet charged only under the new flavour gauge group as a simple dark matter (DM) candidate. This represents a first step in understanding if 
the new flavour gauge bosons are viable candidates as mediators to a dark sector. We find this to be the case, with the relic density obtained via freeze-out in a mostly resonant annihilation regime: the bounds on the gauge coupling constant are compatible with the ones derived from flavour observables.

A broad range of flavourful scenarios has been explored in the literature and provides useful points of comparison for the framework considered in this work. Vector-mediated simplified models with flavour-dependent interactions have been investigated in Refs.~\cite{Goudelis:2023yni,Zhevlakov:2023jzt,Nha:2026pvq}, while low-energy constraints on light flavourful mediators and the implications for charged lepton flavour violation (LFV) have also received considerable attention (see e.g. Refs.~\cite{Farzan:2015hkd,Heeck:2016xkh,Kriewald:2022erk,Davidson:2022nnl,Eguren:2024oov,MartinCamalich:2025srw,Ziegler:2026kis}). Complementary approaches based on simplified scalar mediators with purely LFV couplings have been considered in Ref.~\cite{Ponten:2024grp}, and related phenomenology involving axion-like particles (ALPs) has been explored in Ref.~\cite{Davoudiasl:2021mjy}.
On the experimental side beyond the existing bounds considered in this work, future dedicated facilities targeting muon-to-electron transitions provide a particularly relevant testing ground for these scenarios~\cite{Hill:2023dym,Knapen:2023zgi}, while searches at FASER have also been proposed to probe flavourful new physics~\cite{Araki:2022xqp,Balkin:2024qtf}. A broader overview of current and future experimental programmes can be found in Refs.~\cite{Beacham:2019nyx,Davidson:2022jai}, and complementary signatures involving muon-in-flight conversion have been discussed in Ref.~\cite{Abada:2016vzu}.

The rest of this paper is organised as follows. In Sec.~\ref{sec:models}, we introduce the leptonic flavour-transfer framework and present an explicit example connecting the flavour-transfer operators to the generation of leptonic mass hierarchies.
In Sec.~\ref{sec:constraints}, we study the current laboratory constraints in order to estimate the main classes of available limits. The resulting constraints on the parameter space are presented and discussed in Sec.~\ref{sec:results}, together with possible implications for dark matter scenarios. Finally, we summarise our conclusions in Sec.~\ref{sec:conclusions}. Technical details concerning the implementation of the model and numerical calculations are collected in the appendices.

\section{Models of leptonic flavour transfers }
\label{sec:models}

Flavour transfer is a direct consequence of the presence of a non-abelian flavour gauge symmetry in the UV theory. The resulting gauge bosons can indeed ``transfer'' the flavour between the fermions belonging to a gauge doublet and are thus the primary mediators for tree-level flavourful processes.
The $SU(2)_f$ covariant derivative, including the new gauge group reads:
\begin{equation}
        D_\mu=\partial_\mu+ig_f V^c_{\mu} \, T_c \ ,
\end{equation}
where the new generators are $3\times3$ matrices built from the $SU(2)$ generators depending on the chosen representation for the fermions, and $V_{c} = (V_p, V_m, V_3)$ are the three $SU(2)_f$ gauge bosons.

In the following, we focus on a leptonic $(L)$ model  in which only left- and right-handed leptons transform under the $SU(2)_f$ gauge group. For illustration purposes, we first choose to embed the first two generations of leptons (left and right-handed) in two $SU(2)_f$ doublets. The corresponding generator is then $T_c=\rm{diag}\,(\tau_c,0)$. In the gauge eigenbasis, the new interaction terms then come from the kinetic term of the leptons,
\begin{equation}
\label{eq:UVint}
    L \supset -g_f\,V_\mu^c\, (\bar L_L\gamma^\mu T_cL_L)-g_f\,V_\mu^c\, (\bar E_R\gamma^\mu T_cE_R) \; .
\end{equation}
with $L_L$ an electroweak doublet of left-handed fermions and $E_R$ an electroweak singlet of charged right-handed leptons. It is clear that before specifying the details of the leptonic mass generation, we could have equally embedded the first and third or the second and third lepton generations into a doublet.
In this section, we will explore first the construction of the low-energy models below the electroweak scales based on the above UV structure in Sec.~\ref{sec:spurions}, then focus on a specific flavour model in Sec.~\ref{sec:UVmodel} before discussing in Sec.~\ref{sec:astro} the possible links with astrophysical observables.

\subsection{The emergence of flavour transfer operators}
\label{sec:spurions}

Below the electroweak scale, all the relevant models for flavour-transfer processes must be decomposed into the mass basis.
We define the set of left-handed charged leptons $\ell_L = (e_L, \mu_L, \tau_L)$, of left-handed neutrinos $\nu_L = (\nu_{e_L}, \nu_{\mu_L}, \nu_{\tau_L}) $ and of right-handed charged leptons $\ell_R = (e_R, \mu_R, \tau_R)$.

In order to write the interactions from Eq.~(\ref{eq:UVint}) in the mass eigenbasis, we need to introduce the unitary matrices used in the SM to diagonalise the Yukawa sector. For each fermion type $F$, the diagonalisation matrix can be parametrised as: 
\begin{equation}
    V_F=\Phi V_{12}V_{23}V_{13}\; ,
\end{equation}
where $\Phi=\mathrm{diag}(e^{-i(\phi_1+\phi_2)},e^{i\phi_1},e^{i\phi_2})$ is a phase matrix and the $V_{ij}$ are $SU(3)$ sub-matrices,
\begin{equation}
    V_{23}=\mtx{
        1&0&0\\
        0&e^{i\beta_{23}}\cos\theta_{23}&-e^{-i\alpha_{23}}\sin\theta_{23}\\
        0&e^{i\alpha_{23}}\sin\theta_{23}&e^{-i\beta_{23}}\cos\theta_{23}
    },
\end{equation}
and similarly for $V_{13}$ and $V_{12}$ by rotating the $1-3$ or $1-2$ generations instead. The angles $\theta_{12}$ and $\theta_{13}$ can span the whole $[-\pi,\pi]$ domain, while $\theta_{23}$ is a ``latitude'' angle and can only be chosen within $ [-\pi/2,\pi/2]$. Yet, some of these parameters are not physical, and we can leverage global symmetries to remove them. One can always absorb one global phase per type of fermion, and depending on the chosen scalar sector, use a $U(1)$ or $SU(2)$ global symmetry to absorb either an additional phase or the whole $V_{12}$ factor.\footnote{For a more systematic review of this parametrisation, see e.g the appendix of \cite{Darme:2022uzl}.}

\smallbreak
In this work, we restrict ourselves to real matrices, i.e. $SO(3)$ rotations, for our numerical analysis. The general formulae, however, are presented in a form that allows for complex matrices. In the leptonic scenario, the interaction terms in the mass eigenbasis then read:
\begin{equation}
    \mathcal{L}\supset -g_fV_\mu^c\,(\bar\ell_L Q_L^c\gamma^\mu\ell_L+\bar\nu_LQ_L^c\gamma^\mu\nu_L+\bar\ell_RQ_R^c\gamma^\mu\ell_R)\;,
\end{equation}
with the new effective charges $Q_{L,R}^c$ being defined as
\begin{equation}
    Q^c_L=V_L^\dagger T^c V_L\ ,\quad \, Q^c_R=V_R^\dagger T^c V_R \ .
\end{equation}
Note that, as in the SM, the left-handed charged leptons and the neutrinos share the same diagonalisation matrix. In the following, we will consider a completely degenerate gauge sector, with equal masses for the $V^c$ gauge bosons. We then have:
\begin{equation}
    V_L=V_{12}^LV_{23}^LV_{13}^L\ ,\quad \, V_R=V_{12}^RV_{23}^RV_{13}^R \ ,
\end{equation}
and in the real case, only 6 angles in total are used to parametrise the model (the three $\theta_{12}^L,\theta_{13}^L$ and $\theta_{23}^L$  together with the three $\theta_{12}^R, \theta_{13}^R$ and $\theta_{23}^R$).\footnote{Note that a $SU(2)$ global symmetry can be further used to remove the $V_{12}$ rotation from the right-handed leptons (this choice is arbitrary, and the rotation could equivalently be removed from the left-handed leptons instead).} 
It is of particular interest to study the limit where the diagonalisation matrices are close to identity, $V_L,V_R\sim\mathds1$. Indeed, as no lepton flavour violation (LFV) is observed in the SM, and rather strong bounds are put on possible LFV processes (see below), our model should implement a way to suppress them. This is automatically done by the $SU(2)$ structure of the new interaction, which in the limit of zero mixing angles completely forbids any LFV process and retains only FT processes. Thus, for our model to satisfy these constraints, the mixing matrices $V_L$ and $V_R$ should be close to identity. This will allow us to parametrise all the observables encountered in the subsequent paragraphs in terms of small $SU(2)$-breaking spurions $\theta_{ij}^X$, where $i,j\in\qty{1,2,3}$ and $X\in\qty{L,R}$.

We will explore three main patterns for the spurionic angles:
\begin{itemize}
    \item (FT) Flavour-transfer only - We set all spurions to zero, thus only keeping flavour-transfer observables. This therefore assumes the limit case where the mass eigenstates are entirely aligned with the original gauge eigenstates. We distinguish the three cases of possible leptonic pair included in the $SU(2)_f$ doublets: \FTem~(electron-muon), \FTet~(electron-tau) and \FTmt~(muon-tau).
    \item (FV) All flavour-violating spurions arise proportionally to the ratio of lepton masses $\theta_{ij} = m_i / m_j$, with $i>j$. Since the spurions are small, we still distinguish the three cases of possible leptonic pair included in the $SU(2)_f$ doublets, leading to three models: \FVem~(electron-muon), \FVet~(electron-tau) and \FVmt~(muon-tau).
    \item (UV) We set two angles to order one values  $\theta_{12}^L  \, , \theta_{13}^R \sim \mathcal{O}(1)$ , and $ \theta_{ij} = m_e / m_\tau$ otherwise. This is both motivated by an explicit spurion realisation of the leptonic mass hierarchies described in the section below and also serves as an example of limits in the presence of an order one mixing angle. 
\end{itemize}

\subsection{An example model of leptonic mass hierarchies}
\label{sec:UVmodel}
While the gist of this work focuses on the phenomenological consequences of light new flavourful gauge bosons mediating flavour transfer processes, it is clear that in a complete model of flavour the spurions described above  would arise as predictions linked to the dynamical generation of the lepton mass hierarchies. We present below an example of specific mass spurions which can lead to these hierarchies and derive the corresponding angular spurions introduced in the previous section. 

In general a leptonic $SU(2)_f$ flavour gauge structure is an unpromising starting point for a model of flavour. Indeed, this symmetry structure allows the presence of two Yukawa interactions:
\begin{align}
    \label{eq:Vl}
V_{SU(2)_f} &=y_{12}\, \bar \ell_L^{\,a, i}\, H_{a} e_{R,i} +y_3 \, \bar \ell^{\, a}_{L,3}\,   H_{a} e_{R,3} \ ,
\end{align}
where we have used $a$ indices for $SU(2)_W$, and $i$ for $SU(2)_{f} $. No mass hierarchy is predicted with order one Yukawa interactions. 

We can however relatively easily reverse this conclusion by modifying the UV theory in a way that will not significantly affect the flavour-transfer operators considered in this work. We introduce a supplementary flavour $U(1)_{\lambda_8}$ gauge group (thus moving to $U(2)_f$ instead of $SU(2)_f$), under which all leptons have charge $1$ for the components of the $SU(2)_{f} $ doublet and $-2$ for the $SU(2)$ singlet. Both gauge groups can be straightforwardly embedded in a larger $SU(3)_\ell$ flavour gauge group for further unification attempts. Since $SU(3)_\ell$ is anomaly-free with the SM fermion content supplemented with three right-handed neutrinos, its sub-group is then also anomaly-free with the same fermion content. Equivalently, we can note that on the one hand, the new $U(1)_{\lambda_8}$ is proportional to the well-known anomaly-free combination $2 L_\tau - L_e - L_\mu$ (see e.g.~\cite{Allanach:2018vjg}), and we checked explicitly that mixed flavour-gauge anomalies also cancel out. 

Equipped with this new gauge group, we can prevent the appearance of $SU(2)_f$ conjugate spurions $\varepsilon_{ij} (Y^\dagger)_i$ from a spurion $Y_i$ by charging it under $U(1)_f$, thus allowing for a hierarchy between the two fermionic generations combined in an $SU(2)_f$ doublet. 

However, this still does not prevent the Yukawa structure from Eq.~\eqref{eq:Vl} to appear. We thus go one step further and charge the SM-Higgs under the new $U(1)_f$ flavour gauge group. This will eventually induce a small component of the $U(1)_f$ vector boson to mix at tree-level with the SM $Z$ boson, and thus ultimately induce flavour non-universal interactions to the $Z$. We can however suppress this tree-level mixing by taking a $U(1)_f$ breaking scale large enough, so that we will only study the loop-induced processes in Sec.~\ref{sec:Zboson}.  The new $U(1)_f$ flavour gauge group forbids the operators in Eq.~\eqref{eq:Vl}. We can then re-create the flavour hierarchies by raising the rank of the lepton mass matrix spurion by spurion~\cite{Greljo:2023bix}.
For a concrete example, we take for the $SU(2)_f \times U(1)_f$ charges:
\begin{itemize}
    \item $(\textbf{1},-1)$ for the SM Higgs $H $
    \item $(\textbf{2},2)$ for the first spurion $Y_1^i \propto (y_1, 0)$ generating the third generation mass
    \item $(\textbf{2},4)$ for the second spurion $Y_2^i \propto (y_2, \tilde{y}_2)$ generating the second generation mass
    \item $(\textbf{1},1)$ for the last, $U(1)_f$ only spurion $s$ giving a mass to the first generation.
\end{itemize}
Combining these three spurions we obtain the following structure for the Yukawa sector:
\begin{align}
    \label{eq:Vl2}
V_{SU(2)_f} &=(Y_{1}^\dagger)^i\, \bar \ell_L^{\,a, 3}\, H_{a} e_{R,i} + (Y_{2})^i\, \bar \ell^{\,a, i}_L\, H_{a} e_{R,3} +  s \left( y_{12}\, \bar \ell^{\,a, i}_L\, H_{a} e_{R,i} +y_3  \bar \ell^{\,a}_{L,3}\,   H_{a} e_{R,3}  \right)\ ,
\end{align}
which in turn leads to the mass matrix:
\begin{align}
  \mathcal{M}_e  = v_{EW} \begin{pmatrix}
        y_{12} s & 0 &  y_2\\ 0  & y_{12} s & \tilde{y_2}  \\ y_1 & 0 & y_3 s
    \end{pmatrix} \ .
\end{align}
It is clear that this matrix will generate mass hierarchies in the case we have the following spurions hierarchies $s \ll y_2 \ll y_1$, with a choice roughly fitting the SM spectrum being $s \sim 5\times 10^{-6} \ll y_2, \tilde{y}_2 \sim 5 \times 10^{-4} \ll y_1 \sim 5 \times 10^{-4}$. 
The rotation angles corresponding to these spurion choices are in turn given by:
\begin{align}
    \theta_{12}^L \sim \mathcal{O}(1),  &\quad \theta_{13}^L, \theta_{23}^L \sim \frac{m_e}{m_\tau} \ ,\\
      \theta_{13}^R \sim \mathcal{O}(1), &\quad\theta_{12}^R,  \theta_{23}^R \sim \frac{m_e}{m_\tau} \ .
    \end{align}
Mixing angles of order one arise both in the left-handed and right-handed sectors, however, the mass hierarchies are maintained naturally since the third generation leptonic mass arises mostly from a single component of the left-handed doublet. Similarly, the second generation masses arise from only one component of the right-handed doublet. Since right-handed neutrinos must also be decomposed into an $SU(2)_f$ doublet and singlet, it is clear that the final neutrino mass pattern should eventually depend on the scalar fields generating in the UV the various spurions considered above. We simply point out that in a type-I seesaw 
construction, since we expect the Yukawa neutrino matrix to follow the same hierarchies as the charged lepton one, the large rotations in the PNMS matrix must arise from the $SU(2)_f$-breaking right-handed neutrino Majorana mass term.

\subsection{Astrophysics constraints and dark matter}
\label{sec:astro}

Since horizontal flavour gauge symmetries are a priori unrelated to the other SM gauge groups, their gauge couplings can be in principle significantly smaller, potentially leading to relatively light and feebly-coupled new bosons. Such particles could play a role in the dark matter puzzle and leave various astrophysical signatures. 

First, flavoured gauge can act easily as a new mediator for other dark-sector particles. In order to assess the potential of the above construction in reproducing the observed relic density, we consider the following Lagrangian for the dark sector:
\begin{align}\label{eqn:dm_lagrangian}
\mathcal{L} \supset i \overline{\chi} (\slashed\partial + i g_f \,V^c_\mu \,T_c  \gamma^\mu ) \chi \ +  M_\chi \overline \chi \chi \ ,
\end{align}
with $\chi = (\chi_1,\chi_2)$ a $SU(2)_f$ doublet of Dirac fermions and $SU(2)_f$ indices are left implicit in the above equation. 
 
The dark matter is stabilised in this simple scenario by an accidental dark fermion number
While the Dirac fermion case will be typically subjected to strong constraints below the GeV, both Majorana and Dirac cases can be combined to lead to a pseudo-Dirac dark matter model with a small splitting between both dark matter components (see e.g.~\cite{Tucker-Smith:2001myb,Darme:2017glc,Bell:2021xff,Garcia:2024uwf}).  
The case of a fermionic doublet $\chi$ with small mass splitting, with a dark photon mediator $V$, resonant regime (i.e.~$m_V \approx 2m_\chi$) and $m_\chi \leq 10$ GeV has been studied recently in~\cite{Brahma2024Resonant}. 
In particular, the least stringent limit for $m_\chi = 10$ GeV is $g_\chi \gtrsim 10^{-4}$, while for $m_\chi = 10$ MeV is $g_\chi \gtrsim 10^{-7}$.
In all cases, the dark matter phenomenology will be further subject to flavour constraints, and annihilation will proceed also with mixed flavoured final states (e.g $\bar e \mu , \bar t c, \bar b s$).
As a first order of magnitude estimate, for a freeze-out scenario without a large mediator mass suppression or other exotic behaviour, we expect the relic density target to be of the order
\begin{align}
    \Omega h^2 \sim 0.1 \left( \frac{0.1}{g_f} \right)^4 \,  \left( \frac{m_\chi}{20 \, \textrm{GeV} \times \sqrt{N}}\right)^2 \ ,
\end{align}
with $N$ the number of annihilation channels available, with the relic density of dark matter composed of two $SU(2)_f$-degenerate Dirac fermions. For the scenario under consideration with a flavour gauge symmetry acting on leptons, this implies that the typical couplings realising the adequate relic density are within experimental reach, as we will show in Sec.~\ref{sec:dm_limits}. 
In particular, we focused on the $L_\mu - L_\tau$ case, where the processes contributing to the indirect detection at LO  are at one-loop ($\chi \chi \to \gamma$ via a $\mu$ or $\tau$ loop), so we do not expect them to be very stringent. 
Contrarily, for the baryonic models, we would expect stringent limits coming from direct detection experiments, and -- for indirect detection -- from antiproton fluxes; while the $e^+, e^-$ fluxes are expected to put more stringent limits on the $L_e - L_\mu$ and $L_e - L_\tau$ cases.
Hence, the complete profiling of these models from the dark matter perspective can be the subject of future ad-hoc studies. 
Finally, the case with small mass splitting and resonant annihilation can also be studied, since it would present a different phenomenology \cite{Brahma2024Resonant}.

Numerically, we have leveraged the model implementation in \marty~\cite{uhlrich__2021} described in~\autoref{chap:marty_impl}  loaded in \darkpack\cite{Palmiotto:2022rvw, Palmiotto:2024vpn} and \darksusy\cite{Bringmann:2018lay, Gondolo:2004sc, darksusy_url}, as explained in \autoref{chap:darkpack_darksusy_interface}.

\vspace{0.5cm}

Beyond dark matter, the presence of light flavour gauge bosons can have an impact on several astrophysical observables of interest that we will briefly review below.

First, the light flavour bosons have an unsuppressed gauge interaction with neutrinos. Their impact on the neutrino decoupling temperature has been extensively studied for new abelian $L_i - L_j$ gauge groups (see e.g.~\cite{Escudero:2019gzq} for a complete calculation of the $L_\mu-L_\tau$ case). While a specific study of our model should be performed in order to obtain an accurate prediction for the effective number of neutrino species $\Delta N_{\rm eff}$. We typically expect the corresponding constraint from CMB measurements~\cite{Planck:2018vyg,Escudero:2019gzq} to occur in the $m_V \lesssim 5 - 10$ MeV mass range.\footnote{We note that in this low mass range the flavour constraints will be shown to be in any case so strong -- leading to $g_f \lesssim 10^{-9}$ -- that the corresponding bosons may not necessarily be in thermal equilibrium in the early universe~\cite{Escudero:2019gzq}.} We will in the rest of this article restrict ourselves to gauge bosons heavier than $10$ MeV, thus also simultaneously alleviating potential stellar cooling bounds~\cite{Escudero:2019gzq,Foldenauer:2024cdp,Bauer:2018onh,An:2013yfc}.
In the $10-100$ MeV mass range and for couplings of order $g_f \sim 10^{-9} - 10^{-7}$, it is on the other hand likely that our model would be subjected to bounds from the SN1987 supernova event, as is known for the $L_\mu - L_\tau$ case (see for instance the recent~\cite{Blinov:2025aha}) or for the dark photon type of models (see e.g~\cite{Chang:2018rso}). Since obtaining quantitative results typically requires a complete description of the model, including the possible presence of dark matter, we will however concentrate on accelerator-based searches and leave this study to future works.

Finally, the $SU(2)$ flavour gauge constructions used in this work are known to lead to potential strong gravitational wave signatures for large gauge couplings $g_f$~\cite{Greljo:2019xan,Chrysostomou:2025vrg}. The projected reach from the proposed third-generation gravitational wave observatories such as the EINSTEIN telescope~\cite{Punturo:2010zz} are typically complementary to the accelerator constraints we will evaluate in detail in this work and could help probing in the future our models in the deep UV regime where $g_f \sim 1$ and $m_V \sim 10-1000$ TeV.

\section{Constraints from laboratory experiments}
\label{sec:constraints}

Our goal is to estimate the main classes of limits on the models described above. First, in Sec.~\ref{sec:PureFT} we consider tree-level leptonic flavour transfer processes and all other relevant processes which do not rely on the presence of flavour-violating spurions. Second, we collect instead in Sec.~\ref{sec:LFV} observables which are either loop-induced, or proceed only in the presence of non-negligible spurions.

We used \marty~\cite{uhlrich__2021} to perform the analytical calculation of either the squared decay/scattering amplitudes or effective coefficients. The resulting scalar expressions were converted by \marty~ into a \texttt{C++} library for numerical evaluation. When needed, the phase-space integration of the squared amplitudes has been done using the \texttt{VEGAS} Monte-Carlo algorithm as implemented in the \texttt{GSL} library. More details about the implementation of the model in \marty~ and the numerical phase-space implementation can be found in App.~\ref{chap:marty_impl}.

\subsection{Pure FT processes}
\label{sec:PureFT}

\subsubsection{Neutrinos trident and scatterings }
\label{sec:trident}

Since neutrinos can be easily produced with different flavours as incoming particles and are indistinguishable as final state particles in all relevant processes, they play a key role in constraining our scenarios in the case of negligible flavour spurions. We consider in this section both neutrino trident processes and neutrino-electron scattering experiments.

\paragraph{Neutrinos trident}
One of the best available constraints on neutrino interactions arises from the study of neutrino trident interactions. The interference with the SM process leads to a strong effect up to relatively large masses, as shown in~\cite{Altmannshofer:2014pba}. 

Experimentally, the limit from the CCFR collaboration~\cite{CCFR:1991lpl,Altmannshofer:2014pba} dominates the constraints, although a more recent analysis from the NuTeV collaboration claimed that certain backgrounds were not included~\cite{NuTeV:1999wlw,Greljo:2021npi}. 
\begin{align}
    \frac{\sigma_{\textrm{trid.}}^{CCFR}}{\sigma_{\textrm{trid.}}^{\textrm{SM}}} = 0.82\pm 0.28  \ .
\end{align}

For both of these experimental results, the input neutrino beam consists of $\nu_\mu$ and $\bar{\nu}_\mu$ in the ratio 2:1, as typically obtained from proton beam dump neutrino generations. The average neutrino energy is $\langle E_\nu \rangle = 160 $ GeV~\cite{CCFR:1991lpl} and both CCFR and NuTeV used an iron target.

One specificity of the model we are considering is that several mediators may contribute to the final cross-section, along with various leptonic final states. 

We are thus interested in calculating the cross-section for the process $\nu N\to \nu N\ell^+\ell^-$, where $\ell$ is a SM lepton and $N$ is a nucleus. This cross-section can be approximated using the \textit{equivalent photon approximation} (EPA), which relates it to the cross-section of the process $\nu\gamma\to \nu\ell^+\ell^-$, where $\gamma$ is a real photon. In the real process, a virtual photon with virtuality $q^2$ is emitted by the nucleus and reacts with the passing neutrino. The EPA states that:
\begin{equation}
    \sigma(\nu N\to\nu N\ell^+\ell^-)=\int \sigma(\nu\gamma\to\nu\ell^+\ell^-)\,\dd P(s,q^2)\ ,
\end{equation}
where $s$ is the center-of mass energy of the $\nu\gamma$ system, and the probability density $\dd P(s,q^2)$ reads:
\begin{equation}
    \dd P(s,q^2)=\frac{Z^2e^2}{4\pi^2}\frac{\dd s}{s}\frac{\dd q^2}{q^2}F^2(q^2)\ ,
\end{equation}
where $F(q^2)$ is the electromagnetic form factor of the nucleus.

\begin{figure}[t]
    \centering
    \includegraphics[width=\linewidth]{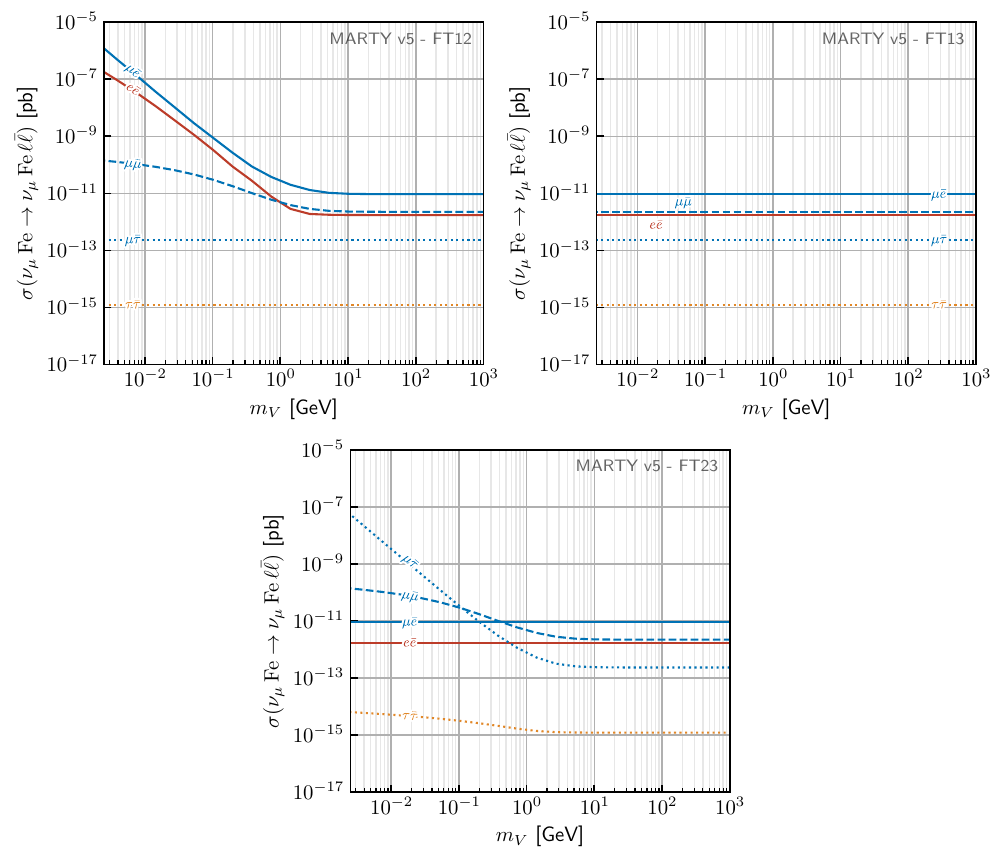}
    \caption{Cross-section for muonic neutrino trident production on iron as a function of the unified flavour gauge boson mass $m_V$  for the different final lepton pairs. We include both SM and NP contributions and study the $SU(2)_f$ flavour-transfer models \FTem, \FTet~and\FTmt. We set in all the plots $g_f=10^{-2}$.}
    \label{fig:sigma_tridents_FT}
\end{figure}

\begin{figure}[t]
    \centering
    \includegraphics[width=\linewidth]{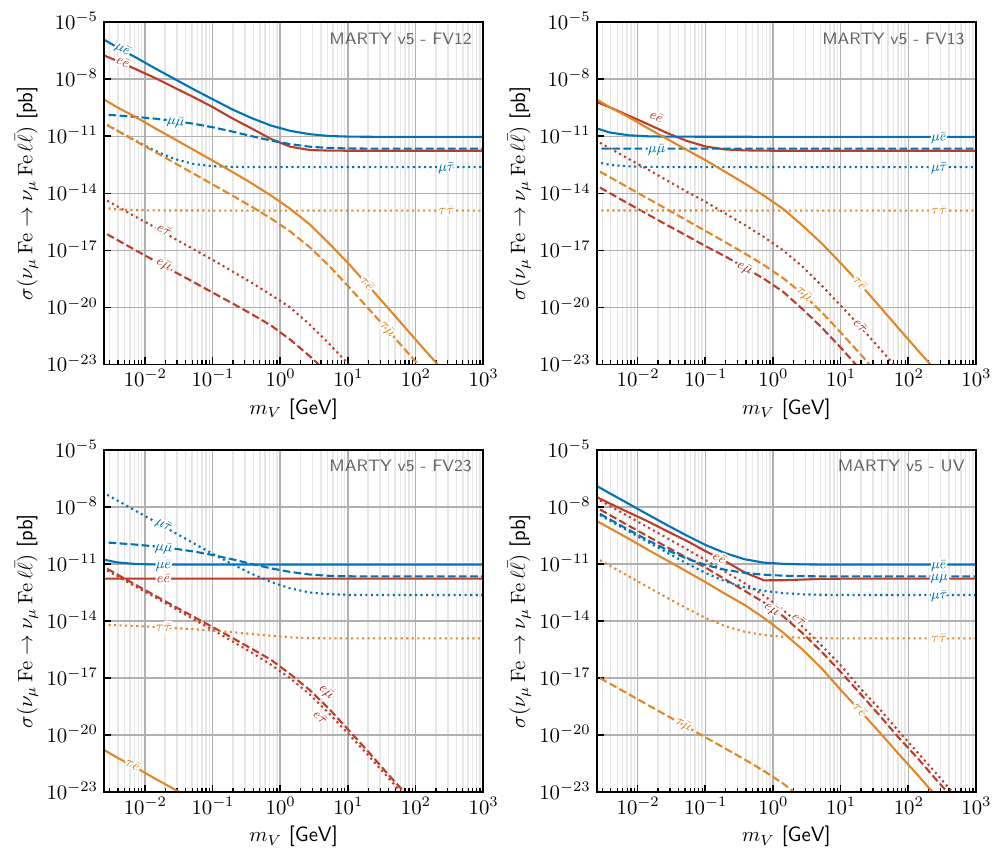}
    \caption{Cross-section for muonic neutrino trident production on iron as a function of the unified flavour gauge boson mass $m_V$  for the different final lepton pairs. We include both SM and NP contributions and study the $SU(2)_f$ flavour-transfer models \FVem, \FVet, \FTmt~and (UV). We set in all the plots  $g_f=10^{-2}$.}
    \label{fig:sigma_tridents_FV}
\end{figure}

We have used \marty\ to generate each individual $\sigma(\nu\gamma\to\nu\ell^+\ell^-)$, tolerating any outgoing neutrino flavour while enforcing an (anti)-muon neutrino as the incoming particle. While all experimental searches are based on a di-muon final state, flavour transfer processes can easily lead to a mixed $e \mu$ or $\tau \mu$ final state. We show in Fig.~\ref{fig:sigma_tridents_FT} the trident cross-sections on an iron target for the different leptonic final states for the three pure flavour-transfer models. We included both the NP and SM contributions to the cross-section and chose a gauge coupling of $g_f = 0.01$. The \FTet~does not lead to any corrections since the flavour gauge bosons do not couple to $\nu_\mu$ in this model. Remarkably, the  \FTem~and\FTmt~models predict the largest deviations in the $\mu e$ final states, in stark contrast with usual NP trident searches for an $L_\mu - L_\tau$ boson. This represents a relatively specific signature of the presence of our NP flavour transfer. We then show in Fig.~\ref{fig:sigma_tridents_FV} the same cross-sections, this time in the presence of flavour-violating spurions. While the cross-sections for most final states are not modified, we note the appearance of several NP-specific final states, including in particular $e \bar{\tau}$ and $\bar{e} \tau$. The (UV) model with large spurions is particularly interesting in this regard.

Note that the future DUNE near-detector may improve the above limit by a factor of around $2$, but only in the low mass region below around $1$ GeV due to the significantly lower neutrino mean energy~\cite{Altmannshofer:2019zhy}.

\paragraph{Neutrino scattering}

We can conclude this section by briefly discussing neutrino scattering experiments, focusing again on the limit of negligible spurions. Due to the absence of tree-level quark interactions, only neutrino-electron scattering processes can be tested. 

The first relevant limits arise from reactor electronic neutrinos with low energy, preventing in practice the production of $\mu$ or $\tau$ leptons in the final state. The dominant NP contribution can therefore only arise from a diagonal flavour process mediated by a $V_3$ gauge boson. Since our model predicts a vectorial interaction, we can use the constraints from reactor electron neutrinos in GEMMA~\cite{Beda:2009kx} and TEXONO~\cite{TEXONO:2009knm} as compiled in~\cite{Lindner:2018kjo}. Since the outgoing neutrinos can be of any flavour, we re-interpret the limit as:
\begin{align}
    g^{SU2}_{lim} = g^{1803}_{lim} \left( \sum_{a=1,2,3}  \left[ \sum_{V=V_3,V_p,V_m} Q^V_{\nu_e, \nu_a} Q^V_{ee} \right]^2\right)^{-1/4} \ ,
\end{align}
where $Q^V_{\nu_e, \nu_a}$ are the various effective charges of our gauge bosons for the final fermionic mass eigenstates as derived from the \marty~numerical subroutines.
We further include the constraints from CHARM-II~\cite{CHARM-II:1994dzw} as reported in~\cite{Lindner:2018kjo}, although they rely instead on muonic neutrinos so that we must make the replacement $Q_{\nu_e, \nu_a} \to Q_{\nu_\mu, \nu_a}$ in the above formula.
In practice, we found these limits to be subdominant in all the relevant parameter space.

\subsubsection{Leptonic invisible decays}
\label{sec:STU}

When considering the decay of a charged lepton $\ell_j$, flavour transfer processes lead to leptonic final states of the form $ l_i \bar{\nu}_i \nu_j$. Since the final state mimics the $W$-induced charged current in the SM, care must be taken when estimating the limits. In particular, the measurement of muon lifetime, while extremely precise, is an input of the SM which serves to fix the bare parameter of the SM Lagrangian. We therefore cannot use the NP-induced process $\mu \to e \bar{\nu}_e \nu_\mu$ directly as a constraint. Instead, we will follow a procedure similar to the one used in defining the oblique parameters $S$, $T$ and $U$~\cite{Peskin:1991sw,Burgess:1993qk}. We first observe that assuming that our NP process would directly impact the muon lifetime measurement, we would obtain a shift
\begin{align}
    \Gamma_\mu^{SM} \to   \Gamma_\mu^{SM} ( 1 + \delta_\mu) \ .
\end{align}
However, since $ \Gamma_\mu^{SM}$ serves as the definition of $G_F$ in the SM, such a change corresponds to a NP contribution to the $T$ parameter of
\begin{align}
    \alpha T = - \delta_\mu/2 \ ,
\end{align}
which can be compared to the latest experimental fit~\cite{ParticleDataGroup:2024cfk} 
\begin{align}
  T_{\rm{exp}} =   0.04 \pm 0.12 \ .
\end{align}
It is further expected that at FCC-ee the precision on this parameter may reach $0.0022$~\cite{deBlas:2019rxi}. This constraint also remains relevant even in the presence of spurions when the $\mu \to e V$ channel is open kinematically but the $V$ bosons are long-lived, such that they escape the experimental apparatus. We also expect the missing energy distribution to be completely different from the one induced by SM charged-currents.\footnote{This could be matched to shifts in the Michel parameters for this decay~\cite{ParticleDataGroup:2024cfk}. Deriving a limit based on this discrepancy would however require a complete recasting of the experimental search since the monochromatic missing energy spectrum shape would strongly depend on the experimental reconstruction capabilities. We will therefore use the above oblique variable result also in this case as a first conservative estimate.}

For $\tau$ invisible decays, we can instead directly rely on the experimental value~\cite{ParticleDataGroup:2024cfk} dominated by the Belle measurement~\cite{Belle:2013teo} since it does not enter the SM renormalisation procedure.
\begin{gather}
 	\tau_\tau = (2.903 \pm 0.005) \cdot 10^{-13}\, s \ . 	  
\end{gather}
The above limit can again be applied in all cases. Note that we could also use the more recent Belle II result~\cite{Belle-II:2024vvr} on the non-universality of leptonic decay
\begin{align}
    R_\mu = 0.9675 \pm 0.0007 \pm 0.0036 \ , 
\end{align}
which should be compared to the predicted SM value $ R_\mu^{SM} = 0.9726$. However, our prediction for this value is strongly spurion-dependent.
We further note that the most recent average of $\tau$ lepton properties leads to a small enhancement of the $\tau$ leptonic decay rates compared to the expectation from the muon lifetime~\cite{HeavyFlavorAveragingGroupHFLAV:2024ctg}, which somehow weakens the corresponding constraint.\footnote{Note that for very light new gauge boson, searches leveraging the monochromatic nature of the outgoing lepton have been suggested for axion-like particles and could be of relevance (see Ref.~\cite{Calibbi:2020jvd}).}

\subsubsection{Accelerator searches }

When the flavour gauge group encompasses first-generation leptons, $e^\pm$-based accelerator searches lead to significant constraints as the $V_3$ gauge boson behaves in practice as an effective dark photon candidate (or more specifically a $L_\mu - L_e$ or $L_\tau - L_e$ gauge boson). We have leveraged the code \texttt{DarkCast}~\cite{Ilten:2018crw,Baruch:2022esd} to recast existing dark photon searches from both beam dump experiment~\cite{Andreas:2012mt,Riordan:1987aw,Bross:1989mp,Konaka:1986cb,Davier:1989wz,Banerjee:2019hmi} and $e^+ e^-$ colliders~\cite{Ablikim:2017aab,Lees:2014xha,Anastasi:2015qla,Anastasi:2016ktq,Anastasi:2018azp}.

When no $e^\pm$ interactions are available at tree-level, we expect that kinetic mixing will generate them at one-loop, although the limits in this case are typically irrelevant compared to the muon and tau decays described in the previous section. In recent years, searches studying the production of muon-philic bosons as final state radiation $\mu^+ \mu^- X,  (X \to \mu^+ \mu^-)$ have steadily improved while relying only on the NP coupling to muons. We have thus included visible searches for $L_\mu - L_\tau$ from CMS~\cite{CMS:2018yxg} and BaBar~\cite{BaBar:2016sci} (note that the Belle-II collaboration is also providing equivalent limits~\cite{Belle-II:2022yaw,Belle-II:2024wtd}).

Finally, the NA64$\mu$ experiment~\cite{NA64:2024nwj} used a muon beam to explore  Bremsstrahlung-like processes of the form $$ \mu Z \to \mu Z X \ ,$$ with $Z$ a nucleus and where the muon-philic new boson is assumed to decay invisibly. The corresponding limits can be straightforwardly applied to the on-shell emission of the $L_\mu - L_\tau$ boson, although it only covers the low mass region where all scenarios are subjected to stronger constraints.

\subsection{Precision constraints and flavour-violation observables} 
\label{sec:LFV}
In general, flavour models will predict a non-zero mixing for the charged leptons sector. This can be traced to the simple fact that each flavour gauge breaking spurion can in general contribute to several entries in the leptonic mass matrix (as shown for instance in our example UV model). We consider thus in this section the processes generated by flavour-breaking operators beyond the flavour-transfer case, leveraging the example scenarios \FVem, \FVet, \FVmt~and (UV) presented in Sec.~\ref{sec:spurions}.

\subsubsection{Rare lepton decays} 

In the limit of heavy $V$ bosons, one can obtain the following effective Lagrangian, containing dipole, anapole and 4-fermion operators,
\begin{equation}
    \mathcal{L}_{\rm eff} \supset F_X^{\alpha\beta}\qty\big(\bar\ell_\alpha\sigma^{\mu\nu}P_X\ell_\beta)F_{\mu\nu}+A_X^{\alpha\beta}\qty\big(\bar\ell_\alpha\gamma_{\mu}P_X\ell_\beta)\partial_\nu F^{\mu\nu}+C_{XY}^{\alpha\beta\gamma\delta}\mathcal{O}_{\alpha\beta\gamma\delta}^{XY} + \rm h.c.
\end{equation}
Where $X,Y\in\qty{L,R}$, $\alpha, \beta, \gamma, \delta$ are generation indices for fermions in the mass basis and 
\begin{equation}
    \mathcal{O}_{\alpha\beta\gamma\delta}^{XY}=\qty\big(\bar{\ell}_\alpha\gamma^\mu P_X\ell_\beta)\qty\big(\bar{\ell}_\gamma\gamma_\mu P_Y \ell_\delta) \ .
\end{equation}
Note that there exists some redundancy in these 4-fermion operators, due to the commuting fermion bilinears and Fierz identities. In particular, 
\begin{equation}
    \mathcal{O}_{\alpha\beta\gamma\delta}^{XX}=\mathcal{O}_{\alpha\delta\gamma\beta}^{XX}=\mathcal{O}_{\gamma\delta\alpha\beta}^{XX}=\mathcal{O}_{\gamma\beta\alpha\delta}^{XX} \ ,\quad \mathcal{O}_{\alpha\beta\gamma\delta}^{X\bar X}=\mathcal{O}_{\gamma\delta\alpha\beta}^{\bar XX} \ ,
\end{equation}
where $\bar X$ is the opposite chirality of $X$ (i.e. if $X=L$ then $\bar X=R$ and vice versa).
In our model, the dipole and anapole operators are generated at the loop level while LFV 4-fermion operators are already generated at tree level, hence we will stay at tree level when studying the latter. Analytical expressions for the loop-generated effective couplings have been studied in the literature, see e.g. \cite{Crivellin:2013hpa, Langacker:2000ju} or \cite{Chiang:2011cv,Kriewald:2022erk,Davidson:2022nnl}, and read:
\begin{gather}\label{eq:A}
    A_X^{\alpha\beta}=\frac{g_f^2}{16\pi^2m_V^2}\sum_{\gamma,c}Q_{X,c}^{\gamma\alpha}\qty\big(Q_{X,c}^{\gamma\beta})^*\frac{6\log(x_\gamma)-1}{9} \ ,\\
    F_X^{\alpha\beta}=\frac{g_f^2 e}{48\pi^2m_V^2}\sum_{\gamma,c} \qty[3  Q_{\bar X,c}^{\alpha\gamma}(Q_{X,c}^{\beta\gamma})^*m_\gamma-Q_{X,c}^{\alpha\gamma} (Q_{X,c}^{\beta\gamma})^*m_\alpha-Q_{\bar X,c}^{\alpha\gamma}(Q_{\bar X,c}^{\beta\gamma})^*m_\beta] \ ,\label{eq:F}
\end{gather}
where $x_\gamma=m_\gamma^2/m_V^2$ and $e$ is the electric charge $e=\sqrt{4\pi\alpha_{\rm em}}$. The couplings for the tree-level generated operators can be readily computed in the redundant basis,
\begin{equation}\label{eq:C}
    C_{XY}^{\alpha\beta\gamma\delta}=\frac{g_f^2}{m_V^2}\sum_c Q_{X,c}^{\alpha\beta}Q_{Y,c}^{\gamma\delta} \ .
\end{equation}
From these couplings, we can express the relevant observables to study the phenomenology of this model. In particular, we will consider the existing limits on the branching fractions of LFV processes including three-lepton and radiative decays, and on anomalous lepton magnetic moments. 

This effective theory is only relevant above the highest mass scale coupling to the new bosons, which is, in our case, the $\tau$ lepton's mass. Hence, in the following we will trust the effective theory above $m_V\gtrsim\SI{10}{\giga\electronvolt}$. In the light boson regime $m_V\lesssim\SI{10}{\giga\electronvolt}$, the $V$ bosons cannot be integrated out and one needs to take their propagation into account, which is done numerically within the~\marty~framework (see App.~\ref{chap:marty_impl}).

\paragraph{Rare three-body lepton decays.}

We start with the study of LFV three-lepton decays of the form $\ell_\alpha\to\ell_\beta\ell_\gamma\bar \ell_\delta$. Current (from SINDRUM \cite{SINDRUM:1987nra} and Belle \cite{Hayasaka:2010np} collaborations) and future (from the Mu3e experiment \cite{arndt_technical_2021} and Belle II \cite{kou_belle_2020} collaboration) experimental limits for the relevant decay channels are gathered in table~\ref{tab:const_lll}. 

The general width for these decays can be computed directly from the effective Lagrangian in the case of heavy-flavour bosons. Depending on $\ell_\beta,\ell_\gamma$ and $\ell_\delta$, different contributions can arise at loop-level (e.g. from $V$ and electroweak penguins), but the dominant contribution for all these decays comes from 4-fermion operators generated at tree-level from the UV theory.
Keeping only tree-level contributions, the decay width reads~\cite{Langacker:2000ju}:
\begin{align}
    \Gamma(\ell_\alpha\to\ell_\beta\ell_\gamma\bar\ell_\delta)=\frac{Sm_\alpha^5}{1536\pi^3}\bigg(\qty\big|C_{LL}^{\beta\alpha\gamma\delta}+C_{LL}^{\gamma\alpha\beta\delta}|^2 + \qty\big|C_{LR}^{\beta\alpha\gamma\delta}|^2 + \qty\big|C_{LR}^{\gamma\alpha\beta\delta}|^2 \\
    \qty\big|C_{RL}^{\beta\alpha\gamma\delta}|^2 + \qty\big|C_{RL}^{\gamma\alpha\beta\delta}|^2 + \qty\big|C_{RR}^{\beta\alpha\gamma\delta}+C_{RR}^{\gamma\alpha\beta\delta}|^2\bigg)\ , \nonumber
\end{align}
where $S$ is a symmetry factor, with $S=1/2$ if two outgoing fermions are identical (such as in $\mu^\pm\to e^\pm e^\pm e^\mp$) and $S=1$ otherwise.
\begin{table}[t]
    \centering
    \begin{tabular}{c|c|c}
    \hline
        Process & Current BR limit & Future BR limit\\\hline
        $\quad\mu\to ee\bar e\quad$ & $<\num{1.0e-12}$ \cite{SINDRUM:1987nra} & $\lesssim 10^{-16}$ \cite{arndt_technical_2021} \\
        $\tau\to ee\bar e$ & $<\num{2.7e-8}$ \cite{Hayasaka:2010np} & $\lesssim\num{5e-10}$ \cite{kou_belle_2020} \\
        $\tau\to e\mu\bar\mu$ & $<\num{2.7e-8}$ \cite{Hayasaka:2010np} & $\lesssim\num{4.5e-10}$ \cite{kou_belle_2020}\\
        $\tau\to\mu\mu\bar e$ & $<\num{1.7e-8}$ \cite{Hayasaka:2010np} & $\lesssim\num{2.5e-10}$ \cite{kou_belle_2020}\\
        $\tau\to\mu e\bar e$ & $<\num{1.8e-8}$ \cite{Hayasaka:2010np} & $\lesssim\num{3e-10}$ \cite{kou_belle_2020}\\
        $\tau\to ee\bar\mu$ & $<\num{1.5e-8}$ \cite{Hayasaka:2010np} & $\lesssim\num{2.2e-10}$ \cite{kou_belle_2020}\\
        $\tau\to\mu\mu\bar\mu$ & $<\num{2.1e-8}$ \cite{Hayasaka:2010np} & $\lesssim\num{3.5e-10}$ \cite{kou_belle_2020}\\\hline
    \end{tabular}
    \caption{Summary of the current and future experimental limits on LFV three-body lepton decay branching fractions.}
    \label{tab:const_lll}
\end{table}

We can separate the processes of interest having two indistinguishable leptons in the final state (of the form $\ell_\alpha\to\ell_\beta\ell_\beta\bar\ell_\gamma$) from the others (of the form $\ell_\alpha\to\ell_\beta\ell_\gamma\bar\ell_\gamma$). In the first case, the expression of the decay width simplifies to
\begin{equation}
    \Gamma(\ell_\alpha\to\ell_\beta\ell_\beta\bar\ell_\gamma)=\frac{m_\alpha^5}{1536\pi^3}\qty(2\qty\big|C_{LL}^{\beta\alpha\beta\gamma}|^2+\qty\big|C_{LR}^{\beta\alpha\beta\gamma}|^2+\qty\big|C_{RL}^{\beta\alpha\beta\gamma}|^2+2\qty\big|C_{RR}^{\beta\alpha\beta\gamma}|^2)\ .
\end{equation}
As an example, in the \FVem~model with parametrically small LFV, one finds the following semi-analytical branching fractions for the relevant processes in terms of the new physics scale $m_V/g_f$,
\begin{equation}\label{eq:spur_nr}
    \mathcal{B}(\ell_\alpha\to\ell_\beta\ell_\gamma\bar\ell_\delta)=10^{-5}\qty(\frac{\SI{1}{\tera\electronvolt}}{m_V/g_f})^4\left\{\begin{array}{ll}
        \num{46}\qty({\theta_{12}^L}^2+{\theta_{12}^R}^2) & (\mu\to ee\bar e) \\
        \num{3.1}\qty({\theta_{13}^L}^2+{\theta_{13}^R}^2) & (\tau\to ee\bar e)\\
        \num{6.2}\qty({\theta_{23}^L}^2+{\theta_{23}^R}^2) & (\tau\to\mu e\bar e)\\
        \num{6.2}\qty({\theta_{13}^L}^2+{\theta_{13}^R}^2) & (\tau\to e\mu\bar\mu)\\
        \num{3.1}\qty({\theta_{23}^L}^2+{\theta_{23}^R}^2) & (\tau\to\mu\mu\bar\mu)
    \end{array}\right.,
\end{equation}
while $\tau\to ee\bar\mu$ and $\tau\to\mu\mu\bar e$ decays lead to weaker $\mathcal{O}({\theta_{ij}^X}^4)$ constraints. 

In the light boson regime, the effective theory breaks down and a complete calculation is performed using \marty~(see details in the appendix). We show in Fig.~\ref{fig:gamma_tau_lept} the ratio of the NP rare $\tau$ leptonic decay width over the total $\tau$ experimental width for the various flavour scenarios considered in this work.

A number of relevant features are worth discussing. 
First, whenever it is kinematically allowed, the on-shell production of the $V$ bosons dominates the decay width. The physical process is then $\ell_\alpha\to \ell_\beta V(\to\ell_\gamma\bar\ell_\delta)$ (or the same with $\beta$ and $\gamma$ swapped). This leads to a decay width of the form 
\begin{equation}
\Gamma(\ell_\alpha\to\ell_\beta\ell_\gamma\bar \ell_\delta)=\Gamma(\ell_\alpha\to\ell_\beta V)\mathcal{B}(V\to\ell_\gamma\bar\ell_\delta)+(\beta\leftrightarrow\gamma)\propto \frac{g_f^2}{m_V^2}\ ,
\end{equation}
The final scaling is in this case in $\frac{g_f^2}{m_V^2}$
instead of the $g_f^4/m_V^4$ scaling of the true three-body decay. This manifests as a jump of several orders of magnitude in the decay width at the mass thresholds where the on-shell channels open or close, and a different slope in log-scale between the opening and closing thresholds. For a given decay, the mass thresholds are:
\begin{equation}
    \min(m_\beta+m_\delta,m_\gamma+m_\delta)<m_V<\max(m_\alpha-m_\beta,m_\alpha-m_\gamma) \ .
\end{equation}
For $\tau$ decays, the upper threshold is always approximately $m_\tau$, but depending on the exact decay channel, the lower threshold can either be around $2m_\mu,m_\mu$ or $2m_e$. In some cases, there are two distinct sub-processes for a given decay channel, e.g. $\tau^-\to\mu^-V(\to e^-e^+)$ and $\tau^-\to e^- V(\to \mu^-e^+)$, with different mass thresholds. This manifests as two separate jumps at the corresponding thresholds. These are especially visible with hierarchical spurions as the size of the jump is proportional to the coupling of the $V$ boson to each lepton pair. A good example of this behaviour is seen in Fig.~\ref{fig:gamma_tau_lept} in the \FVmt~model, where the $\tau e$ and $\mu e$ couplings are much more suppressed than the $\tau\mu$ coupling, resulting in large jumps at the muon and di-muon thresholds. At very low mass, below the on-shell thresholds, the leading contribution arises from the pseudoscalar Goldstone boson coupling to the leptons, corresponding to the longitudinal polarisation of the $V$ bosons (see appendix on the implementation of the model in \marty), yielding back a scaling as $g_f^4/m_V^4$. One important feature of this regime is the pseudoscalar nature of the coupling, leading to an amplitude proportional to the difference of the lepton's masses. For models with vectorial couplings, i.e. in which $\theta_{ij}^L=\theta_{ij}^R$, the Goldstone bosons do not couple to pairs of same-flavour leptons, hence the saturation of the $\tau\to\mu\mu\mu$ decay width in the (FV) models which are indeed vectorial in their couplings. The same behaviour would be observed for the $\tau\to eee$ decay width below the di-electron threshold. 

\begin{figure}[t]
    \centering
    \includegraphics[width=\linewidth]{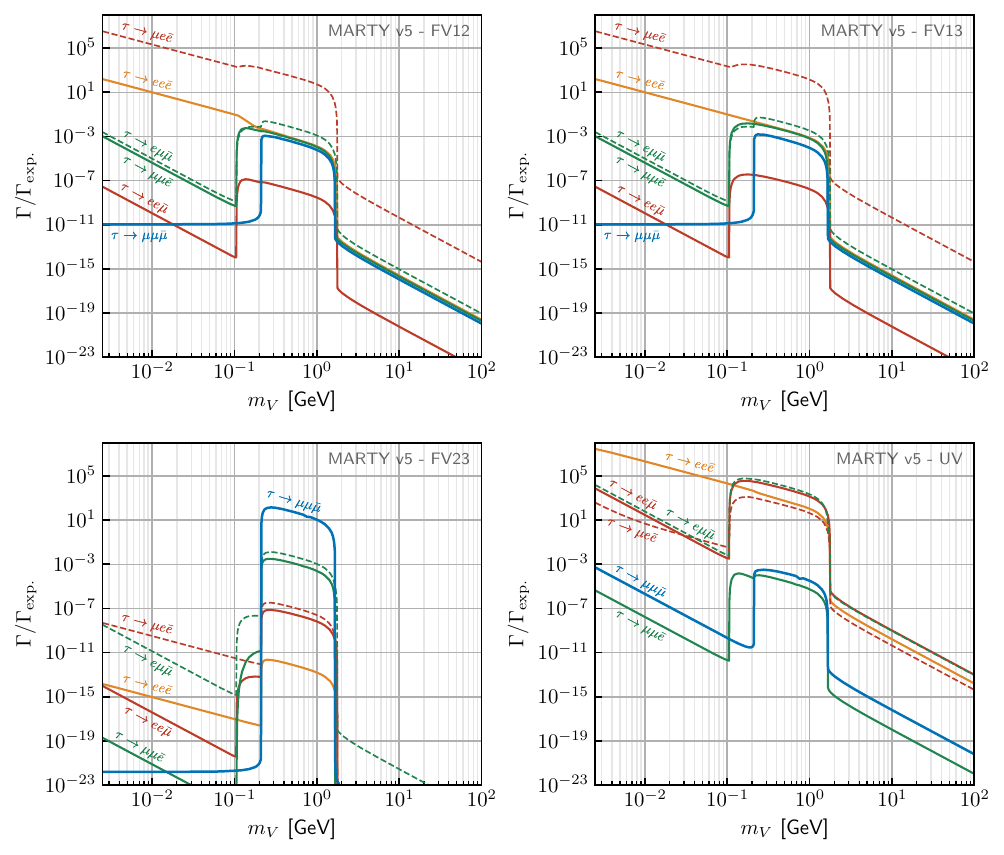}
    \caption{  Ratio of the partial $\tau$ decay width into flavour-violating leptonic over the total $\tau$ experimental width as a function of the unified flavour gauge boson mass $m_V$. Dashed lines denote decays with no identical lepton pair in the final state ($\tau^+\to\mu^+e^+e^-$ and $\tau\to\mu^+e^+\mu^-$). We include  the $SU(2)_f$ flavour-violating models \FVem, \FVet, \FTmt~and (UV). We set in all the plots  $g_f=10^{-3}$.}
    \label{fig:gamma_tau_lept}
\end{figure}

\begin{figure}[t]
    \centering
    \includegraphics[width=\linewidth]{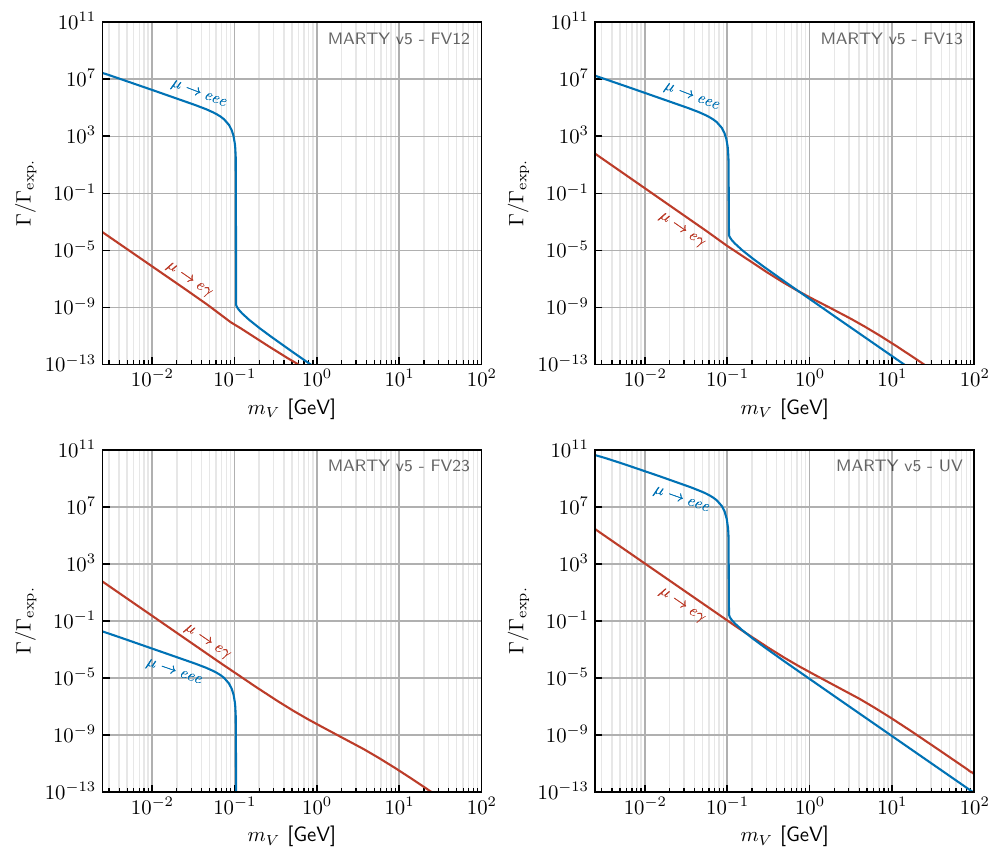}
    \caption{Ratio of the partial $\mu$ decay width into flavour-violating leptonic $eee$ and $e\gamma$ final states over the total $\mu$ experimental width as a function of the unified flavour gauge boson mass $m_V$. We show the $SU(2)_f$ flavour-violating models \FVem, \FVet, \FTmt~and (UV). We set in all the plots  $g_f=10^{-3}$.}
    \label{fig:gamma_mu}
\end{figure}

The rare muonic decay $\mu\to eee$ is shown in Fig.~\ref{fig:gamma_mu}. The same behaviour as in the $\tau$ decays is observed, only with the upper on-shell threshold being at $m_\mu$ instead of $m_\tau$. Here it is clear how the $SU(2)$-breaking spurions affect the decay width: in the models with hierarchical spurions $\theta_{ij}\sim m_i/m_j$, the \FVem~and \FVet~models yield similar decay widths while the \FVmt~model yields a dramatically lower decay width. Indeed, in the \FVem~model the dominant contribution in the on-shell regime will come either from $\mu\to eV_3(\to ee)$ or $\mu\to eV_m(\to ee)$ with a $\theta_{e\mu}^2$ breaking factor. In the \FVet~model, the dominant contribution will come this time only from $\mu\to eV_3(\to ee)$, still with a $\theta_{e\mu}^2$ breaking factor. Comparatively, in the \FVmt~model, the dominant contribution still comes from $\mu\to eV_3(\to ee)$, but this time with a $\theta_{e\mu}^6$ breaking factor as each electron needs to be ``rotated'' into a muon in order to couple to the $V_3$ boson, explaining the $m_e^4/m_\mu^4\sim10^{-10}$ order of magnitude difference with the other two models.

\paragraph{Radiative lepton decays.}

Radiative leptonic decays of the form $\ell_\alpha\to\ell_\beta\gamma$ are also well-constrained \cite{MEG:2016leq, BaBar:2009hkt, uno_search_2021}, with increasing precision yet to come from the Belle II experiment \cite{kou_belle_2020}. Relevant experimental limits are summarised in Tab.~\ref{tab:const_ly}.

The partial decay width for these processes in the heavy flavour boson limits reads~\cite{Crivellin:2013hpa, Chiang:2011cv}:
\begin{equation}
    \Gamma(\ell_\alpha\to\ell_\beta\gamma)=\frac{(m_\alpha^2-m_\beta^2)^3}{4\pi m_\alpha^3}\qty(\qty\big|F_L^{\alpha\beta}|^2+\qty\big|F_R^{\alpha\beta}|^2)\approx\frac{m_\alpha^3}{4\pi}\qty(\qty\big|F_L^{\alpha\beta}|^2+\qty\big|F_R^{\alpha\beta}|^2) \ ,
\end{equation}
as kinematics impose that $\ell_\beta$ be lighter than $\ell_\alpha$, thus $m_\beta^2\ll m_\alpha^2$. In the small spurions limit for the model \FVem, one finds for example the corresponding branching ratios:
\begin{equation}\label{eq:spur_r}
    \mathcal{B}(\ell_\alpha\to\ell_\beta\gamma)=10^{-7}\qty(\frac{\SI{1}{\tera\electronvolt}}{m_V/g_f})^4\left\{\begin{array}{ll}
        \num{63}\qty({\theta_{12}^L}^2+{\theta_{12}^R}^2) & (\mu\to e\gamma) \\
        \num{1.4}\qty({\theta_{13}^L}^2+{\theta_{13}^R}^2) & (\tau\to e\gamma)\\
        \num{1.4}\qty({\theta_{23}^L}^2+{\theta_{23}^R}^2) & (\tau\to\mu\gamma)
    \end{array}\right.,
\end{equation}
with this time all observables putting constraints at $\mathcal{O}({\theta_{ij}^X}^2)$. From this and \eqref{eq:spur_nr}, we see that the strongest constraints on muon physics will primarily impact the value of the mixing angle $\theta_{12}^L$ between the first two generations, while the angles $\theta_{13}^X$ and $\theta_{23}^X$ are less strongly constrained by $\tau$ physics limits. 
\begin{table}[t]
    \centering
    \begin{tabular}{c|c|c}
    \hline
        Process & Current BR limit & Future BR limit\\\hline
        $\quad\mu\to e\gamma\quad$ & $<\num{4.2e-13}$ \cite{MEG:2016leq} & -- \\
        $\tau\to e\gamma$ & $<\num{3.3e-8}$ \cite{BaBar:2009hkt} & $\lesssim\num{1e-9}$ \cite{kou_belle_2020} \\
        $\tau\to \mu\gamma$ & $<\num{4.2e-8}$ \cite{uno_search_2021} & $\lesssim\num{3e-9}$ \cite{kou_belle_2020}\\\hline
    \end{tabular} 
    \caption{Summary of the current and future experimental limits on LFV radiative lepton decay branching fractions.}
    \label{tab:const_ly}
\end{table}

In the light boson regime, we used \marty\ to compute the full decay width $\ell_\alpha\to\ell_\beta\gamma$ at one loop.
Predicting which channel is dominant in each model is not as straightforward as in the tree-level purely leptonic decays, as there are several interfering diagrams with different scaling on the lepton mass ratios that can counteract the effect of spurions. For example, the couplings of the $V$ bosons to the leptons favour different flavours to run in the loop depending on the particular model, while the mass insertion needed to flip the chirality of the running lepton always favours the $\tau$ sometimes leading to compensation of the spurionic effects. 

\begin{figure}[t]
    \centering
    \includegraphics[width=\linewidth]{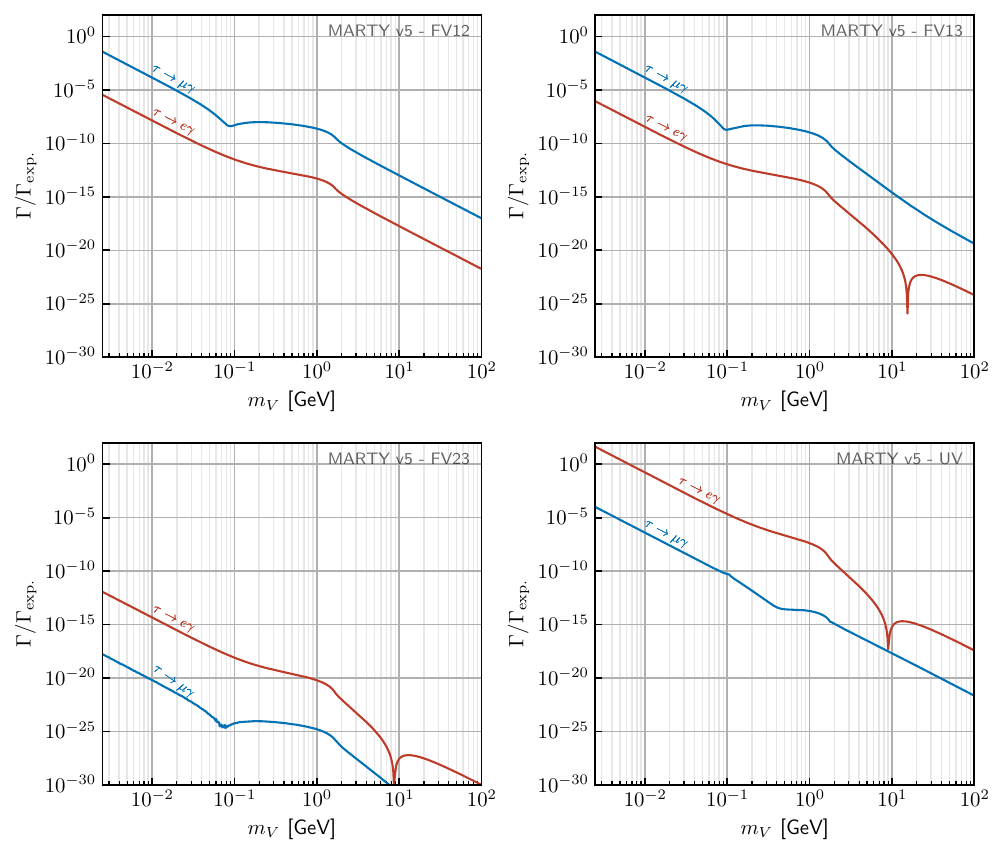}
    \caption{Ratio of the partial $\tau$ decay width into flavour-violating $\mu\gamma$ and $e\gamma$ final states over the total $\tau$ experimental width as a function of the unified flavour gauge boson mass $m_V$. We show the $SU(2)_f$ flavour-violating models \FVem, \FVet, \FTmt~and (UV). We set in all the plots  $g_f=10^{-3}$.}
    \label{fig:gamma_tau_rad}
\end{figure}

We conclude this section with some comments on the corresponding limits that can be obtained from the processes described above.
First, while we chose to normalise all figures to a gauge coupling of $g_f = 0.001$ (or $g_f = 0.01$ depending on the process), it is quite clear that for lower masses, the actual coupling should be several orders of magnitude larger in order to pass experimental constraints. Fortunately, rescaling the results shown in Figs.~\ref{fig:gamma_tau_lept},\ref{fig:gamma_mu} and \ref{fig:gamma_tau_rad} is straightforward since, as discussed in the previous sections, the NP contribution to the decay width scales as $g_f^4$ with the only exception of the region in which the on shell processes are allowed in between mass-threshold, where the decay width scales instead as $g_f^2$ (for the loop-induced radiative decays one also recover a $g_f^4$ scaling).
Second, while on-shell $V$ production processes lead to the most stringent limits, they are subjected to the additional constraint that the produced flavour gauge boson must actually decay promptly inside of the experimental apparatus (otherwise it would instead mimic to an extent a standard electroweak-driven decay with neutrinos in the final state). We will discuss this last point in more detail in Sec.~\ref{sec:results} and in App.~\ref{chap:longlifetime}.

\subsubsection{Lepton magnetic moments}

\smallbreak
Within our effective theory, the lepton anomalous magnetic moment and electric dipole moment read \cite{Moroi:1995yh}:
\begin{gather}\label{eq:a_l}
    a_\alpha=\frac{4m_\alpha}{e}\Re(F_L^{\alpha\alpha}+F_R^{\alpha\alpha}) \ ,\\
    d_\alpha=2\Im(F_L^{\alpha\alpha}-F_R^{\alpha\alpha}) \ ,
\end{gather}
where $\alpha$ is not summed over in the right-hand side. From these expressions, we can immediately see that the lepton electric dipole moment (EDM) is non-zero only if the new gauge bosons couple differently to left-handed and right-handed leptons and if the couplings have a non-zero imaginary part. Therefore, in our simplified model with real rotation matrices, the EDM analysis is irrelevant, as all Wilson coefficients are real. In the limit of small spurions for the \FVem~model and keeping only the non-constant contributions from $m_\tau$, we obtained the following expressions for the anomalous magnetic moments: 
\begin{gather}
\label{eq:aell}
    a_e^{\rm NP}=\frac{m_e m_\mu}{4\pi^2 (m_V/g_f)^2}\qty[1+\frac{m_\tau}{2m_\mu}\qty( \theta_{13}^L \theta_{13}^R+2\theta_{23}^L\theta_{23}^R)] \ ,\\
    a_\mu^{\rm NP}=-\frac{m_\mu^2}{8\pi^2 (m_V/g_f)^2}\qty[1-\frac{m_\tau}{m_\mu}\qty(2 \theta_{13}^L \theta_{13}^R+ \theta_{13}^L \theta_{13}^R)] \ ,\\
    a_\tau^{\rm NP}=-\frac{m_\tau^2}{8\pi^2 (m_V/g_f)^2}\qty[{ \theta_{13}^L}^2+{ \theta_{13}^R}^2+{ \theta_{23}^L}^2+{ \theta_{23}^R}^2] \ .
\end{gather}
As expected, even in the case of negligible spurions, this model predicts a non-zero contribution to the electron and muon magnetic moment. \smallbreak

\begin{table}[t]
    \centering
    \begin{tabular}{c|c|c}
    \hline
        Lepton & $a_\alpha^{\rm NP}$ & $|d_\alpha^{\rm}|$ ($e\,$cm)\\\hline
        $e$ (Cs) & \num{-87(36)e-14} \cite{Fan:2022eto} & $<\num{8.7e-29}$ \cite{ACME:2013pal} \\
        $e$ (Rb) & \num{48(30)e-14} \cite{Fan:2022eto} & $<\num{8.7e-29}$ \cite{ACME:2013pal} \\
        $\mu$ & \num{38(63)e-11} \cite{Aliberti:2025beg} & $<\num{1.9e-19}$ \cite{Muong-2:2008ebm} \\
        $\tau$ & \num{-1(3)e-3} \cite{Pich:2013lsa} & $<\num{5.1e-17}$ \cite{Pich:2013lsa} \\\hline
    \end{tabular}
    \caption{Summary of the current status of the experimental measurements of lepton anomalous magnetic moments and electric dipole moment. }
    \label{tab:const_gm2}
\end{table}

\begin{figure}[t]
    \centering
    \includegraphics[width=\linewidth]{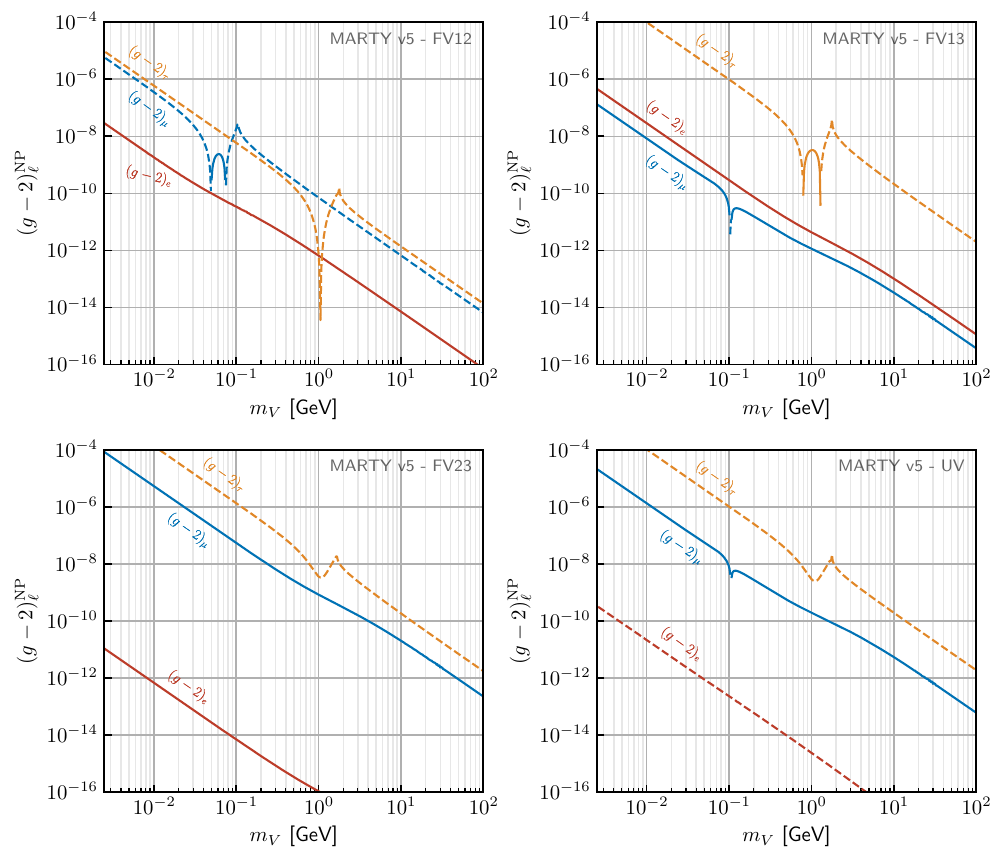}
    \caption{NP contribution to the lepton anomalous magnetic moments $a_e ~\equiv~ (g-2)^{\rm NP}_e, a_\mu~\equiv~ (g-2)^{\rm NP}_\mu$ and $a_\tau  ~\equiv~ (g-2)^{\rm NP}_\tau$ as a function of the unified flavour gauge boson mass $m_V$ in the presence of additional flavour spurions for the \FVem~(top left), \FVet~(top right), \FVmt~(bottom left) and (UV) (bottom right) models. Dashed lines represent negative values and plain lines positive values. We set in all the plots  $g_f=10^{-3}$. }
    \label{fig:g-2lep}
\end{figure}

We have used \marty\ to determine directly the full effective parameters $F_L$ and $F_R$ in each of our benchmark models. The corresponding lepton anomalous magnetic moments for a reference value of $ g_f=10^{-3} $ are shown in Fig.~\ref{fig:g-2lep} for $\tau$ (orange line), $\mu$ (blue line) and $e$ (red line), and the filled (resp. dashed) lines represent positive (resp. negative) values. As could be inferred from the effective theory approach, the loop with $\tau$ lepton induced by $\Wt$ tends to dominate the results, with non-conserved Goldstone currents diverging at small $m_V$, in a clear distinction with the usual $L_{\ell_i}-L_{\ell_j}$ scenarios. The interplay between the $\Zt$ and $\Wt$ further leads to several exact localised cancellations which are however both $m_V$ and $g_f$ dependent.

The anomalous magnetic moments of leptons are historically among the most scrutinised leptonic observables.
Yet, the current situation is rather unclear, with disagreeing values of the theoretical calculation's inputs between data-driven and lattice QCD determinations. 
For the muon, the most precise experimental determination comes from the FNAL Muon $g-2$ experiment \cite{Muong-2:2025xyk} combined with the BNL result \cite{Bennett:2006fi}. The SM estimate presents on the other hand significant internal discrepancies originating mostly from the $e^+ e^-$ data-driven and lattice-driven evaluations of the hadronic vacuum polarisation contribution~\cite{Aliberti:2025beg}.
For the electron, the SM determination depends on the precise value of the fine structure constant $\alpha_{\rm em}$. This value has been measured in two ways, from the study of Caesium-133 atoms in a matter-wave interferometer \cite{Parker:2018vye,Fan:2022eto}, and Rubidium-87 atoms \cite{Morel:2020dww}, resulting in $\gtrsim 5 \sigma$ discrepant values.
The current measurement of the $\tau$ magnetic moment \cite{Pich:2013lsa} is much less precise than the other two, yet we include it in our set of observables for the sake of completeness.

Overall, this experimental situation (summarised in table~\ref{tab:const_gm2}) makes it complex to set definite bounds. In the following, we have therefore decided to indicate two guidelines for the $a_\ell$ in our plots, corresponding to:
\begin{align}
     |a_e^{\rm NP}| &\simeq   10^{-12} \ ,\\
     |a_\mu^{\rm NP}| &\simeq  10^{-9} \ ,
\end{align}
corresponding to the order of magnitude of the difference between both estimates of $a_e$ and twice the error on the residual discrepancy between the SM prediction and the experimental value of $a_\mu$~\cite{Aliberti:2025beg}. We stress that in both cases, these lines should be interpreted with care: they represent the typical regions in which we expect our models to have an impact on the anomalous magnetic moments and not stringent constraints.

\subsubsection{Radiative mixing to SM gauge bosons}
\label{sec:Zboson}
The $SU(2)_f$ gauge symmetry prevents the appearance of a renormalisable kinetic mixing term between the new gauge bosons and the SM $SU(2) \times U(1)_Y$ ones. However, after the breaking of both gauge symmetries (which we assumed is dominantly encompassed by the mass spurions in our case and diagonal masses for the $SU(2)_f$ gauge boson), loop-induced contributions lead to a non-zero $A- Z-\Zt-\Wt$ mass and kinetic mixing since the new flavour gauge bosons natively interact with SM particles. At this point, it is important to note that in the absence of spurions, the $\Wt$ bosons will never mix with the SM neutral gauge boson due to the residual global $U(1)_f$ flavour symmetry (which in this case amounts to the observation that only a single type of fermions can flow in the loops). 

The one-loop off-diagonal self-energies relating two vector bosons are well-known and can be decomposed  into the transverse and longitudinal pieces:
\begin{align}
\Pi_{V V'}^{\mu \nu} ~\equiv~ (g^{\mu \nu} p^2 - p^\mu p^\nu) \kappa_{V V'} +  g^{\mu \nu}  \Sigma_{V V'}\ ,
\end{align}
where $V = Z, A$ and $V' = V_3, V_p, V_m$. The first piece corresponds to the so-called kinetic mixing and the second to a loop-level off-diagonal contribution in the mass matrix.

Let us focus first on the kinetic mixing with the photon. Each flavour gauge boson will inherit a small interaction with the electromagnetic current, as in the much studied case of dark photon models, which can then be used directly to re-interpret the standard dark photon limit into our formalism. Due to the radiative suppression of this mechanism, this coupling could be relevant only for $SU(2)_f$ constructions favouring second and third leptonic generations. Additionally, in the limit of negligible spurions, the situation perfectly mimics that of $L_\mu - L_\tau$ gauge boson where the loop suppression typically renders electron-driven searches irrelevant (see e.g~\cite{Bauer:2018onh}). We thus do not consider these terms further.

We then consider the mixing with the SM $Z$-boson, which can proceed directly via mass mixing and be enhanced in the limit $M_Z \sim m_V$. First, the kinetic mixing can be absorbed at first order by a re-definition of the gauge fields:
\begin{align*}
Z_\mu \to Z_\mu  - \kappa_{Z V_i} V_i \ .
\end{align*}
We obtain off-diagonal mass matrix components of the form $ \Delta M_{ZV_i} = \Sigma_{V V'} +  \kappa_{Z V_i} M_Z^2 \ ,$ which at first order leads to a mixing angle
\begin{align*}
\theta_{Z V_i} &= \frac{ \left(\Sigma_{Z V_i}-M_Z^2 \kappa_{Z,V_i}\right)}{M_Z^2 -m_{V}^2} \ .
\end{align*}
When $M_Z \sim m_V$ however, the $Z$ width should prevent this mixing from diverging (as $\Gamma_Z \neq \Gamma_V$ in all of the relevant parameter space), we thus chose to saturate the limit when $|M_Z - m_V |\sim \Gamma_Z$, although this is only relevant for a very narrow part of the parameter space. This mixing angle has in principle two main phenomenological consequences: (1) shifting the $Z$-boson mass, (2) allowing the $Z$ boson to inherit a fraction of the $SU(2)_f$ current, leading to flavour non-universal or flavour-violating $Z$ decays.
In practice however, and while this mixing can be sizeable in generic $Z^\prime$ models (see e.g~\cite{Dobrescu:2021vak}), two phenomena conspire in our case to strongly suppress these angles. First, the mass mixing is proportional to the product of the $Z$ and $V$ axial coupling and thus vanishes exactly in the absence of spurions due to the vectorial nature of our new gauge interaction. Second, while non-zero, the kinetic mixing is generated via $\tau$, $\mu$ or $e$ loops, but in the absence of spurions only two flavours contribute with opposite couplings. This implies that the radiatively generated kinetic mixing at the $Z$-scale is suppressed by a factor $m_{\ell_i}^2 / M_Z^2 - m_{\ell_j}^2/M_Z^2$ and will not significantly contribute.

Instead, we have therefore used \marty\ to directly generate the one-loop finite triangular diagrams leading to $Z \to \ell_i^+ \ell_j^-$ decays which do not suffer from the above suppression (and no $m_Z \sim m_V$ enhancement), with a $V$ t-channel exchange leading to either non-universal final states, or LFV spurions leading to a LFV final state.

\begin{figure}[t]
    \centering
    \includegraphics[width=\linewidth]{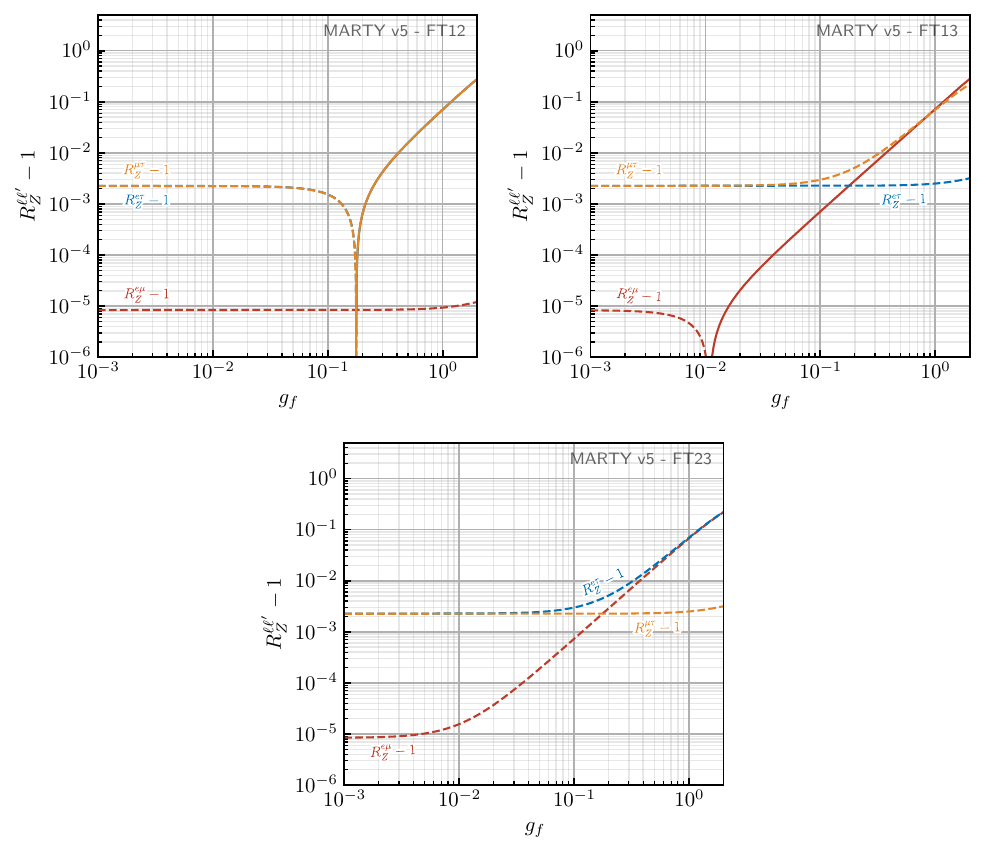}
    \caption{Lepton flavour universality ratios $R_{\ell \ell^\prime} = \Gamma_{Z \to \ell\ell} /  \Gamma_{Z \to \ell^\prime\ell^\prime } $ of the $Z$ boson decays as a function of the gauge coupling $g_f$ in the various flavour transfer models \FTem, \FTet~and \FTmt. Dashed lines are used to represent negative values and plain lines positive values. We set in all the plots  $m_V=\SI{20}{\giga\electronvolt}$.}
    \label{fig:rm1Zllft}
\end{figure}

\begin{figure}[t]
    \centering
    \includegraphics[width=\linewidth]{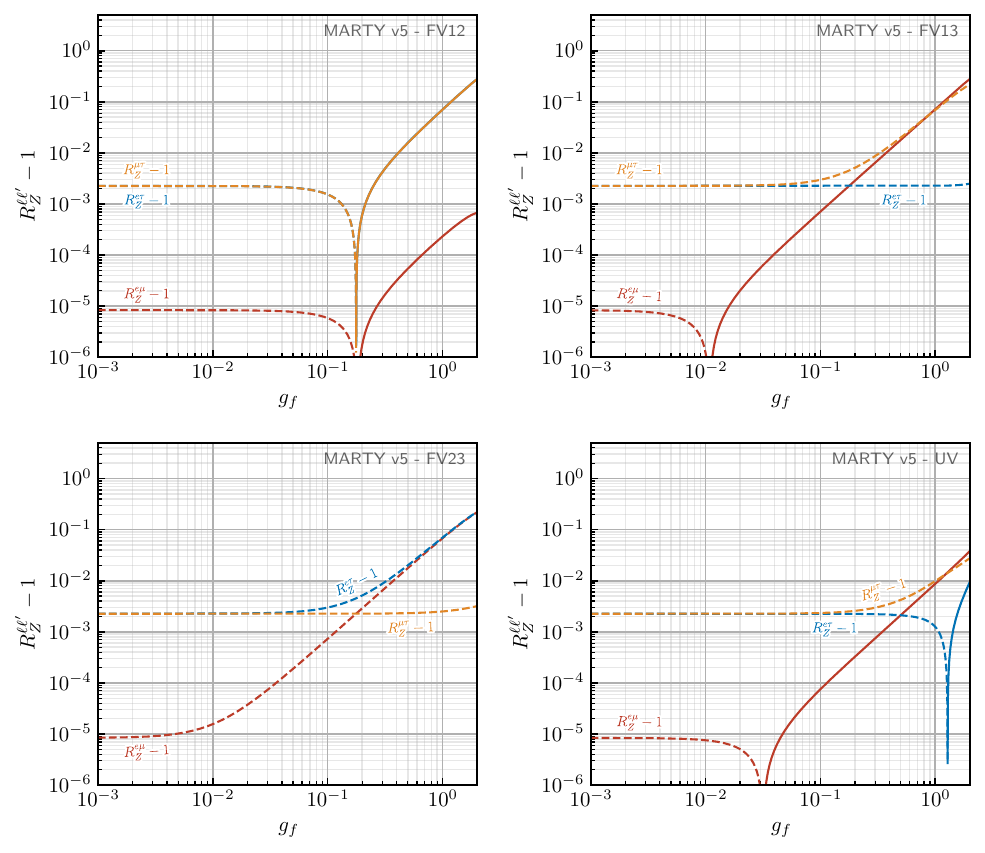}
    \caption{Lepton flavour universality ratios $R_{\ell \ell^\prime} = \Gamma_{Z \to \ell\ell} /  \Gamma_{Z \to \ell^\prime\ell^\prime } $ of the $Z$ boson decays  as a function of the gauge coupling $g_f$ in the various flavour-violating models \FVem, \FVet, \FVmt~and (UV). Dashed lines are used to represent negative values and plain lines positive values.  We set in all the plots $m_V=\SI{20}{\giga\electronvolt}$.}
    \label{fig:rm1Zllfv}
\end{figure}

\begin{figure}[t]
    \centering
    \includegraphics[width=\linewidth]{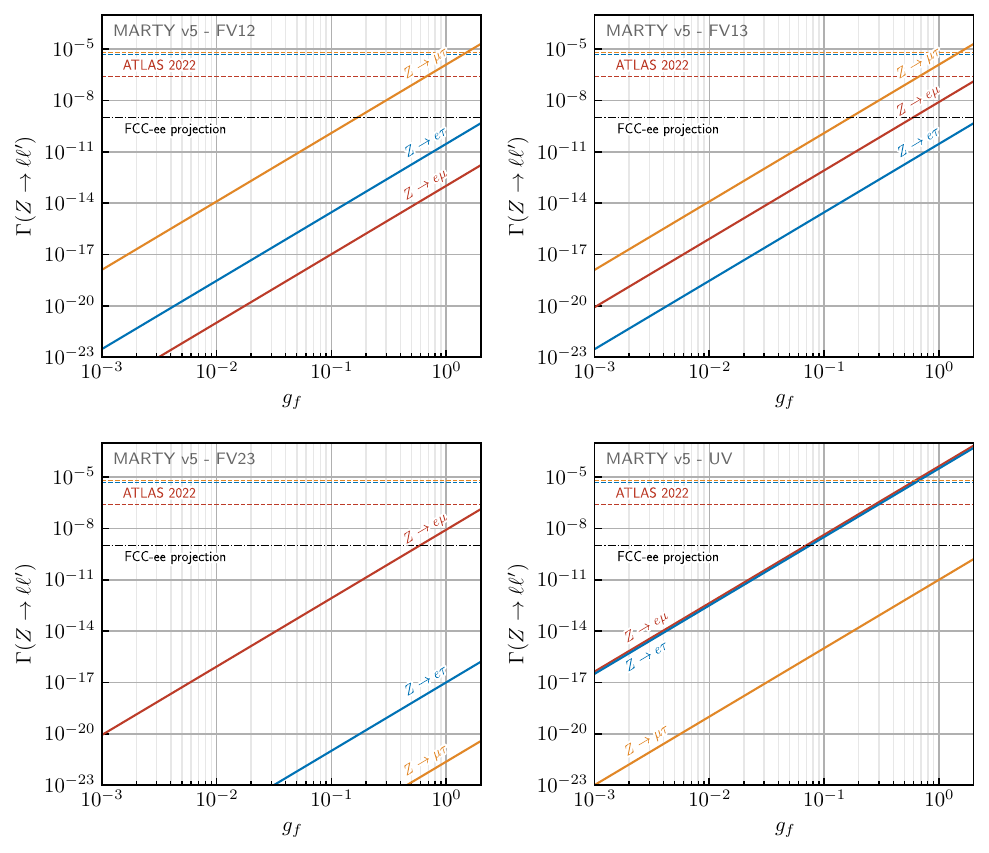}
    \caption{Lepton flavour violating $Z$ boson partial decay width as a function of the gauge coupling $g_f$ in the various flavour violating models \FVem, \FVet, \FVmt~and (UV). In all the plots we take as a reference value $m_V=\SI{20}{\giga\electronvolt}$.}
    \label{fig:gammaZllfv}
\end{figure}

On the experimental side, in the absence of spurions, only the mixing with the $V_3$ gauge boson is allowed, which then generates non-flavour universal interactions for the $Z$. We use the PDG averaged values~\cite{ParticleDataGroup:2024cfk}:
\begin{align}
  &  \Gamma_{Z \to \mu\mu} /  \Gamma_{Z \to ee }  ~=~ 1.0001 \pm 0.0024  \quad \textrm{    (ATLAS~\cite{ATLAS:2016nqi}, LEP average~\cite{ALEPH:2005ab})} \ , \nonumber \\
   & \Gamma_{Z \to \tau\tau} /  \Gamma_{Z \to ee }  ~=~ 1.0010 \pm 0.0026 \quad \textrm{    (LHCb~\cite{LHCb:2018ogb}, LEP average~\cite{ALEPH:2005ab}))} \ , \\
    &\Gamma_{Z \to \tau\tau} /  \Gamma_{Z \to \mu\mu }  ~=~ 1.0020 \pm 0.0032 \quad \textrm{    (LHCb~\cite{LHCb:2018ogb}, LEP average~\cite{ALEPH:2005ab}))} \nonumber \ . 
\end{align}
In the future, a potential FCC-ee machine in its Tera-Z phase could dramatically improve on the precision of these measurements. As rough estimate, We have used in our projection $BR_{Z \to \rm{LFV}} < 10^{-9}$ and $\delta \Gamma_{Z \to \ell_i\ell_i} /  \Gamma_{Z \to \ell_j\ell_j } \lesssim 5 \cdot 10^{-6}$~\cite{Altmann:2025feg}.

We have further considered $Z$ flavour-violating decays which have been strongly constrained by the ATLAS collaboration:
\begin{align}
  &  BR_{Z \to e\mu} ~\lesssim~ 2.6 \cdot 10^{-7} \quad  \textrm{    (ATLAS~\cite{ATLAS:2022uhq})} \ , \nonumber \\
  &  BR_{Z \to e\tau} ~\lesssim~ 5.0 \cdot 10^{-6} \quad  \textrm{   (ATLAS~\cite{ATLAS:2021bdj})} \ , \\
  &  BR_{Z \to \mu \tau} ~\lesssim~ 6.5 \cdot 10^{-6} \quad  \textrm{   (ATLAS~\cite{ATLAS:2021bdj})} \nonumber \ . 
\end{align}

We show in Fig.~\ref{fig:rm1Zllft} the lepton flavour universality ratios of the $Z$ boson decays $R_{\ell \ell^\prime} = \Gamma_{Z \to \ell\ell} /  \Gamma_{Z \to \ell^\prime\ell^\prime } $ in the various flavour transfer models for a fixed $m_V = 20$ GeV, varying instead $g_f$. We include the SM tree-level $Z$ couplings so that the ratio saturates to the kinematically-expected SM values at small $g_f$. Compared to the above constraints, it is clear that we do not expect this limit to overcome other flavour-driven limits. We note however that these processes are typically more sensitive to the details of the full UV theory due to the large $Z$ mass so that this conclusion could be modified in specific models. The dramatic improvements planned at FCC-ee on these measurements will however allow couplings at the few $10^{-3}$ to be probed, allowing new parameter space to be tested via these processes. This conclusion remains for the case of the flavour-violating scenarios shown in  Fig.~\ref{fig:rm1Zllfv}.

We finally show in Fig.~\ref{fig:gammaZllfv} the lepton flavour violating $Z$ boson decay widths in the various (FV) models, along with the corresponding experimental constraints. We fix again $m_V = 20$ GeV and vary instead $g_f$. The absence of corresponding SM processes to interfere with restricts the reach of these bounds. Future FCC-ee measurements in the Tera-Z phase could lead also in this case to an order of magnitude improvement.

\section{Results}
\label{sec:results}

We combine in this last section the various constraints considered previously to offer a state-of-the-art status of the bounds on new $SU(2)$-leptonic flavour gauge bosons, focusing particularly on the lower mass regimes where on-shell effects must be properly included. 

\subsection{Flavour-aligned models}

\begin{figure}[t]
    \centering
    \includegraphics[width=\linewidth]{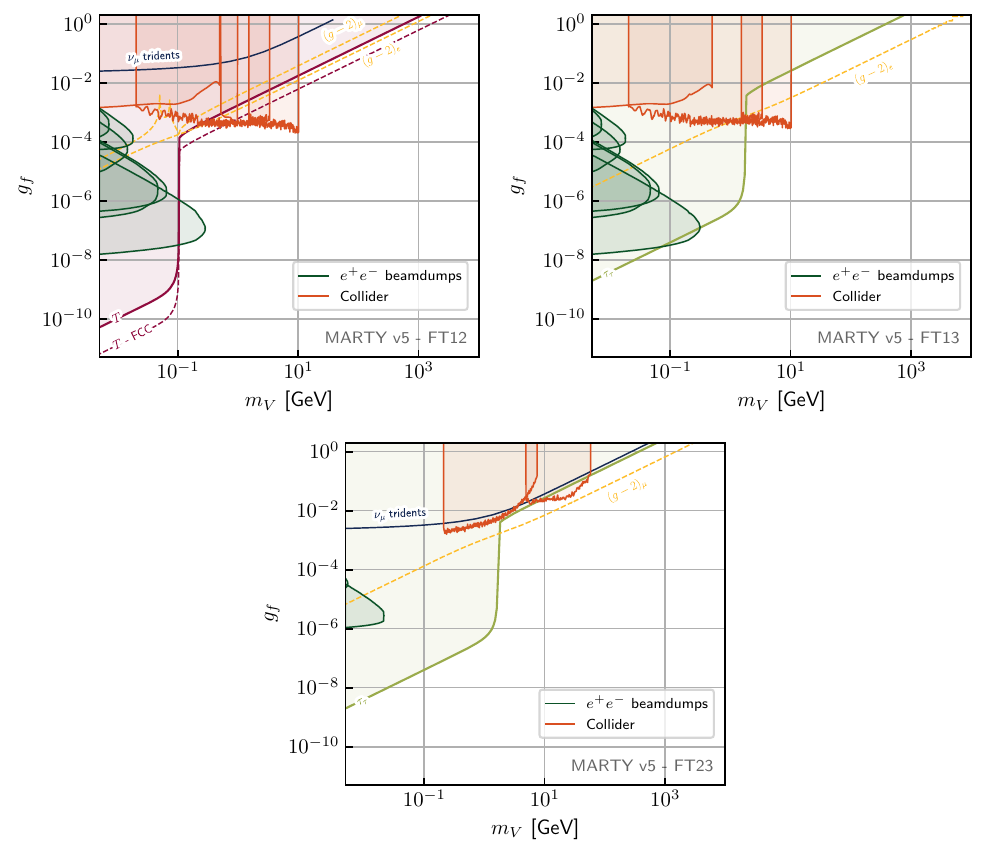}
    \caption{Limits on the flavour coupling $g_f$ as a function of the unified flavour gauge boson mass $m_V$ for the  various flavour transfer models \FTem, \FTet~and \FTmt. We have included muon-coupling driven collider searches (orange area in \FTmt) from CMS~\cite{CMS:2018yxg} and BaBar~\cite{BaBar:2016sci} and electron-driven collider searches~\cite{Ablikim:2017aab,Lees:2014xha,Anastasi:2015qla,Anastasi:2016ktq,Anastasi:2018azp} (orange area in \FTem{} and \FTet{} models); beam dump searches~\cite{Andreas:2012mt,Riordan:1987aw,Bross:1989mp,Konaka:1986cb,Davier:1989wz,Banerjee:2019hmi} are shown in dark green; constraints from $\mu$ (dark red line labelled $T$ with the dashed line showing the FCC-ee projection)~\cite{ParticleDataGroup:2024cfk} and $\tau$ (green line)~\cite{Belle:2013teo} charged decay width; neutrinos trident measurements (purple line)~\cite{CCFR:1991lpl}. An indicative electron $|\delta a_e| \sim 10^{-12} $ and muon $|\delta a_\mu| \sim 10^{-9} $ line is further shown in the plot in dashed yellow.}
    \label{fig:LimitsFT}
\end{figure}

We will focus first on flavour-aligned models \FTem, \FTet~and \FTmt~where there are no flavour-breaking spurions so that only the flavour-transfer or flavour non-universal processes may be used to test the models.
We show in Fig.~\ref{fig:LimitsFT} the summary of constraints along with some projections for each of the three relevant scenarios as a function of the unified flavour gauge boson mass $m_V$.

When interactions with first-generation leptons are available, standard dark photon accelerator-based direct searches can be re-interpreted and provide the dominant constraints around the $1-10$ GeV scale. Below and above this scale, we found the most relevant exclusion to arise from the modifications to the muon or tau lepton lifetimes. This corresponds to the appearance of an oblique parameter $T$ in the \FTem~model constrained by global EW fits in the\FTem~model, and to a direct experimental constraint in the \FTet~model. The only exception is at the tip of the beam dump constraints, typically around $\mathcal{O} (100)$ MeV. Neutrinos trident are found to be subdominant in the entire parameter range. We note that in case constraints from the $e \mu$ final states could be obtained, they would likely help in improving these constraints. The phenomenologically-relevant region of $|a_\mu| \sim 10^{-9}$ is also completely covered in the case of the \FTem~model. On the other hand, the phenomenologically-relevant target for $|a_e|$ of $10^{-12}$ is still not covered by other bounds in both the \FTem~and \FTet~models. In fact, given the strong enhancements of this observable in the latter, it is likely that it could lead to the strongest constraint at large masses once the experimental situation has settled down.

For the \FTmt~scenario, we found that direct experimental searches for $L_\mu-L_\tau$ gauge boson are contributing to the limits again around the $1-10$ GeV scale. At low mass, the $\tau$-lifetime constraint strongly dominates, driven by the on-shell $\tau \to \mu V$ process. Compared to the standard $L_\mu - L_\tau$ case (which corresponds in the absence of spurions to isolating our $V_3$ boson), we find the neutrino trident constraints to be of the same order as the $\tau$-lifetime constraints at large $m_V$ mass. We further note that the muon anomalous magnetic moment is strongly enhanced by $\tau$-driven loops, implying that values of order $ a_\mu \sim 10^{-9}$ are still experimentally allowed above the $\tau$-lepton mass and likely within future experimental reach of Belle-II.

\subsection{Flavour-violating models}

\begin{figure}[t]
    \centering
    \includegraphics[width=\linewidth]{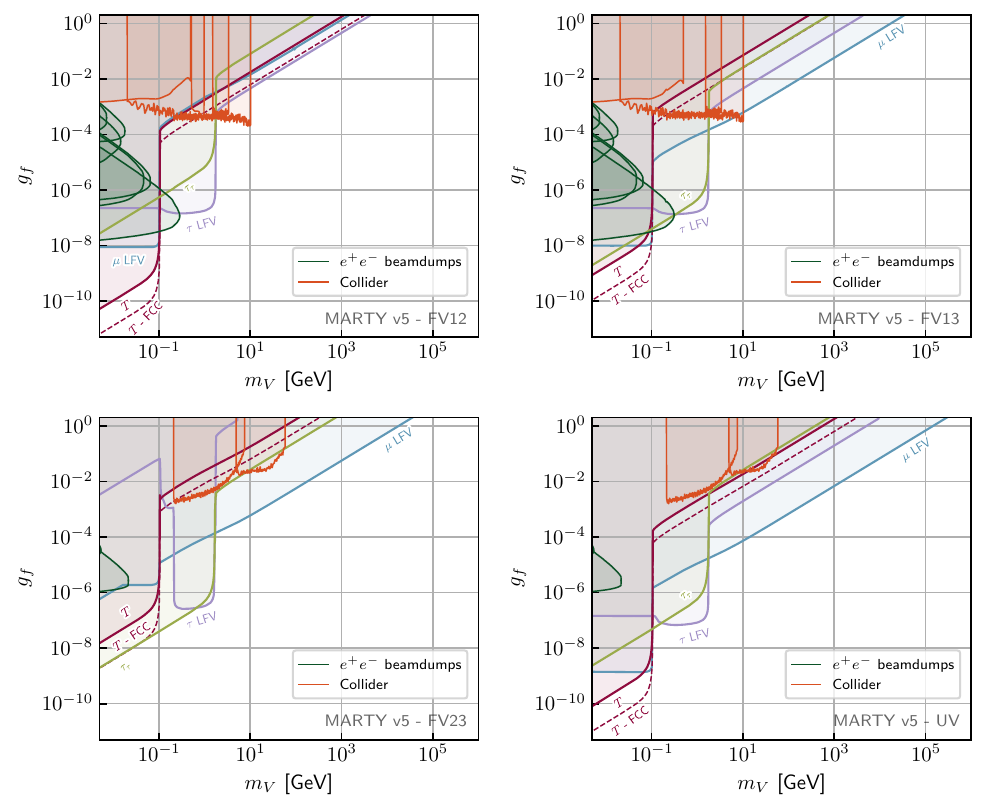}
    \caption{Limits on the flavour coupling $g_f$ as a function of the unified flavour gauge boson mass $m_V$ in the presence of additional flavour spurions for the flavour-violating models \FVem, \FVet, \FTmt~and (UV). We have included muon-coupling driven collider searches (orange area in \FVmt{} and UV) from CMS~\cite{CMS:2018yxg} and BaBar~\cite{BaBar:2016sci} and electron-driven collider searches~\cite{Ablikim:2017aab,Lees:2014xha,Anastasi:2015qla,Anastasi:2016ktq,Anastasi:2018azp} (orange area in \FVem{} and \FVet{} models); beam dump searches~\cite{Andreas:2012mt,Riordan:1987aw,Bross:1989mp,Konaka:1986cb,Davier:1989wz,Banerjee:2019hmi} are shown in dark green; constraints from $\mu$ (dark red line labelled $T$)~\cite{ParticleDataGroup:2024cfk} and $\tau$ (green line)~\cite{Belle:2013teo} charged decay width. Constraints from rare LFV muonic lepton decays $\mu \to e \gamma$, $\mu \to eee$ are shown in blue~\cite{SINDRUM:1987nra,MEG:2016leq} and from $\tau \to \mu/e \ \gamma$, $\tau \to \ell \ell \ell$ in light purple~\cite{BaBar:2009hkt,uno_search_2021,Hayasaka:2010np}.}
    \label{fig:limitsFV}
\end{figure}

We consider in a second time the inclusion of flavour-violating spurions in our theory as described in Sec.~\ref{sec:spurions}. We show in Fig.~\ref{fig:limitsFV} the summary of constraints along with some projections for each of the three relevant scenarios as a function of the new gauge boson mass $m_V$.

In the presence of spurions, rare LFV leptonic decays become available. While they dominate the constraints at large $m_V$ mass, it is remarkable that pure flavour-transfer processes (specifically $\tau$ and $\mu$ lifetime constraints) still represent the strongest exclusions at low mass. This follows from the difficulty in observing a visible decay of $V$ bosons for couplings $g_f \lesssim 10^{-7}$ at small masses due to their long life-time. We give more details regarding our phenomenological treatment of this case in App.~\ref{chap:longlifetime}. For larger coupling, these rare leptons decay processes have a strong mass dependence due to the interplay between the off-shell processes driven by non-conserved currents and the on-shell processes limited by various mass thresholds. 

In the \FVem{} model, the strongest constraint in the high-mass regime is interestingly coming from the $\tau$ LFV decays, despite muonic constraints being significantly more stringent. This reflects the fact that the rare LFV muon decay $\mu\to e\gamma$ which tends to dominate in the muon sector at high $m_V$ in other models is suppressed in this particular flavor alignment. Indeed, the $\tau$ mass loop enhancement is at least suppressed by two spurion factors instead of one in the other alignments. On the other hand, in all the other models, $\tau$ loops are allowed at first order in the spurions, leading to the $\mu$ LFV decays dominating the constraints at high $m_V$.

In all cases the phenomenologically-relevant region of the $(g-2)_\ell$ are completely excluded and are therefore not included in fig. \ref{fig:limitsFV} for the sake of clarity. Finally, we note that the NP scale generating the spurions (corresponding the large $g_f\sim1$, large $m_V$ part of our plots) is constrained to be in the multi-TeV range for the \FVem~model, tens of TeV for the \FVet~and \FVmt~models and in the hundreds of TeV for our UV construction with some order-one spurions.

\subsection{Flavour-transfer $SU(2)_\ell$ as DM mediators} \label{sec:dm_limits}

\begin{figure}
    \centering
    \includegraphics[width=\linewidth]{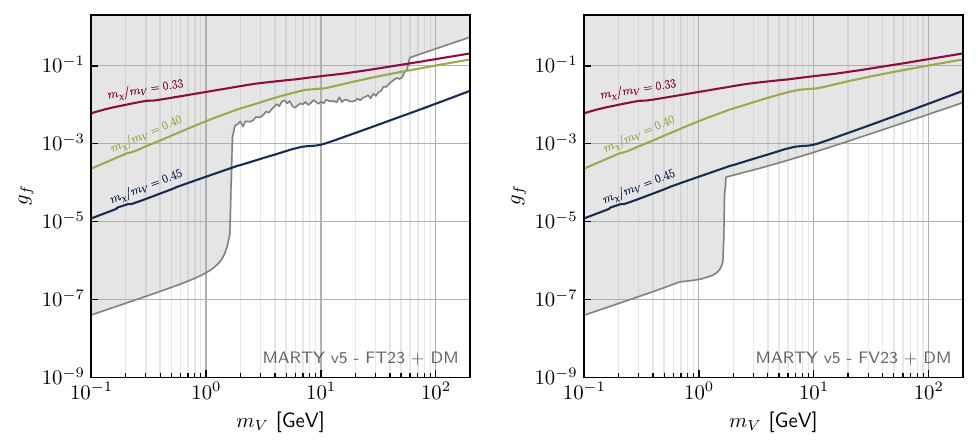}
    \caption{Relic density targets and limits in $g_f$ as a function of the unified flavour gauge boson mass $m_V$. We show  relic density lines corresponding to $\Omega h^2=0.119$ for $m_\chi/m_V = 0.3, 0.4$ and $0.45$. We use the \FTem~ model in determining the DM relic density. The existing bounds considered in this work for flavour-transfer (left plot) and flavour-violating (right plot) models are summarised in grey. 
    }
    \label{fig:relic_density}
\end{figure}

For the dark matter limits, we consider the model with a mass-degenerate Dirac fermion doublet ($\chi_1, \chi_2$), interacting as in Eq.~(\ref{eqn:dm_lagrangian}). We focus on the \FTmt~scenario, but it is clear that apart from the precise location of mass thresholds in the annihilation cross-section, we do not expect the relic density targets to be significantly modified in the \FTet~and \FTem~scenarios.
In the \FTmt~scenario, the dark matter candidate does not directly interact with electrons or quarks; hence there are no sizeable constraints related to indirect detection or direct detection, unlike for the relic density.
We reduce the parameter space to be 2-dimensional 
by fixing the ratio $m_\chi/m_V$ to three different values $0.3, 0.4$ and $0.45$. 
Therefore, the only free parameters left are $g_f$ and $m_V$.\footnote{We assume that any new Higgs scalars breaking the $SU(2)_f$ symmetry are heavier than the $V_a$ and do not directly contribute to relic density calculation.
}

\autoref{fig:relic_density} shows, as a function of $m_V$, the value of $g_f$ required to reproduce the correct relic density in the freeze-out scenario  for the three different values of $m_\chi/m_V$. We focus the mass window around and below the EW scale where most of our limits are concentrated.
As expected, in the context of relic density, the least stringent limit on $g_f$ comes from the case where the annihilation is the closest to the $s$-channel resonance, i.e.~$m_\chi/m_V=0.45$. While a certain level of $s$-channel resonance is required to obtain a thermal target outside the current pure flavour-transfer bounds, it is clear that in this case, the flavour gauge boson can act as an effective DM freeze-out mediator. As illustrated further in Fig.~\ref{fig:relic_density}, including spurions tends to, on the other hand, cover most of the parameter space and would typically imply a dark matter overabundance in the allowed parameter of our DM simplified model.

Furthermore, we note that depending on the mass and coupling regimes, two scalings occur for our relic density targets: $g_f \propto m_V^{1/2}$ and $g_f \propto m_V$ (assuming  $m_\chi/m_V $ is constant) corresponding respectively to an ``off-shell'' or ``on-shell'' $V$ boson exchanges. 
The latter dominates at smaller $g_f$ and when the dark matter masses are closer to the resonance (for the ratio $0.45$). 
Finally, we notice the opening of the different annihilation channels $\mu\mu$, $\mu\tau$ and $\tau\tau$ for values $m_V$ approximately corresponding to the required minimum centre-of-mass energy: depending on the mass ratio, this corresponds to $300-400$ MeV for $\mu\mu$, $3-4$ GeV for $\mu\tau$ and $5-8$ GeV for $\tau\tau$. 
A study on the case of small mass splitting could also be carried out after a future upgrade of the Boltzmann-solver in \darkpack, since \darkpack provides features for computing the averaged annihilation cross-section at low temperature \cite{Arbey:2023yca}.

\section{Conclusions}
\label{sec:conclusions}
In this work, we have studied leptonic flavour-transfer processes arising from an $SU(2)_f$ flavour gauge symmetry. We considered both the flavour-aligned limit, where only flavour-transfer observables are present, and several benchmark flavour-breaking scenarios motivated by leptonic mass generation.

Using a dedicated implementation in \marty, we evaluated the relevant laboratory constraints over a broad range of mediator masses. In the flavour-aligned case, we found that charged-lepton lifetime measurements provide some of the most stringent constraints on the parameter space, while accelerator-based searches and neutrino observables can provide complementary sensitivity depending on the flavour structure. In particular, for the  \FTmt~scenario, values of order $|a_\mu| \sim 10^{-9}$ remain compatible with current constraints in part of the parameter space, while future experimentally-consistent assessments of $a_e$ could significantly constraint the high-mass regime of the \FTet~model. The dominance of charged lepton lifetime bounds likely indicates that the constraints could be improved by a dedicated study of these processes, focusing in particular on the deformation of the SM final state distribution due to the different operators mediating the decay between on-shell (or off-shell) flavour gauge boson emissions and the usual SM charged current.

We then investigated the impact of flavour-breaking spurions. In this case, rare lepton-flavour-violating decays dominate the constraints at large mediator masses. Nevertheless, pure flavour-transfer observables, in particular those derived from muon and tau lifetime measurements, remain the leading exclusions in the low-mass region due to the long lifetime of the flavour gauge bosons which prevent the actual detection of their leptonic decay products.

Finally, we briefly explored the possibility that flavour gauge bosons act as mediators between the Standard Model and a dark sector by using \darkpack. Within the simplified scenario considered here, we found that the observed relic density can be reproduced for couplings and masses lying in the parameter space explored in this work.
A more complete investigation of the dark-matter phenomenology and of scenarios with small mass splittings is left for future studies.

\acknowledgments
The authors thank A.~Deandrea, S.~Monath and F.~Nortier for many interesting discussions. 
M.P.~would like to thank A.~Arbey for the discussions on the Boltzmann-solver, and T.~Bringmann for the help with implementing plugins for \darksusy. The part of the work made by M.P.~has been funded by the Maria Skłodowska-Curie grant agreement No.~101126636, and in a small part by a Universit\'e Claude Bernard Lyon 1 AAP Accueil EC 2024. L.D.~has been supported during the early part of this work by the European Union’s Horizon 2020 research and innovation programme under the Marie Skłodowska-Curie grant agreement No 101028626 up to 31.08.2023.

\newpage

\begin{appendices}

\section{MARTY implementation}\label{chap:marty_impl}
\subsection{Model building in MARTY}
In \marty~\cite{uhlrich__2021}, the $SU(2)_\ell$ models were implemented starting from the existing implementation of the SM (\texttt{mty::SM\_Model}). The new gauge sector was added using \marty's standard features (a comprehensive introduction to \marty's features can be found in its manual \cite{uhlrich2020mtymanual}). The gauge-lepton interactions were added manually in the electroweak-broken phase as it was not straightforward to separate the different lepton generations before the electroweak symmetry breaking which is performed automatically in \marty's SM implementation. Then, as no rigorous scalar sector was built in the UV of our theory, one needs to reconstruct the interaction of the Goldstone bosons $\varphi^a$ associated with the new $V^a$ gauge bosons with the leptons, that would occur as a result of the spontaneous symmetry breaking of our new scalar sector. To retrieve the correct Goldstone-lepton interaction, one can write the BSM part of the Lagrangian as:
\begin{equation}
    \mathcal L\supset -\frac{1}{4}(V_{\mu\nu}^a)^2+\frac{1}{2}m_V^2(V_\mu^a)^2+g_f V_\mu^a \bar\psi\gamma^\mu(Q_L^aP_L+Q_R^aP_R)\psi+\bar\psi M\psi \ ,
\end{equation}
where $V_\mu^a$ are the new gauge bosons with mass $m_V$, $\psi=(\psi_i)$ are chiral fermions, whose left and right-handed components may couple differently to the new gauge group, with masses $m_i$. In the mass basis, $M_{ij}=m_i\delta_{ij}$.

From this expression, we can extract the Feynman rules for the Goldstone-fermion interaction which must compensate the gauge transformation of the mass Lagrangian to restore gauge invariance :
\begin{equation}
    (\varphi^a\bar\psi_i\psi_j) : \frac{g_f}{m_V}\qty[\qty(m_iQ_{L,a}^{ij}-m_jQ_{R,a}^{ij})P_L-\qty(m_jQ_{L,a}^{ij}-m_iQ_{R,a}^{ij})P_R] \ .
\end{equation}
These terms were added explicitly in \marty\ to restore gauge invariance. In practice, for numerical stability, tree-level calculations such as $\ell\to\ell\ell\ell$, $\ell\to\ell\nu\nu$ and $\nu\gamma\to\nu\ell\ell$ were performed in the unitary gauge for the $V$ bosons, and loop calculations such as dipole coefficients, $\ell\to\ell\gamma$ and $Z\to\ell\ell$ were performed in Feynman gauge with Goldstone-lepton interactions. \smallbreak\noindent
The complete \marty~model files can be found as ancillary files to this paper.

\subsection{Numerical integration}
\paragraph{Integration within \marty's numerical library}
The final output of the analytical side of \marty\ is the squared amplitudes and effective coefficients. If one wants to compute decay widths or scattering cross-sections, then the integration over phase-space should be carried out numerically. In order to simplify this process, we developed a small helper suite within the numerical library generated by \marty. The two main objects of this suite are the \texttt{Kinematics} and \texttt{Integrator} objects. \texttt{Kinematics} provides the kinematic inputs needed by \marty, i.e., the $s_{ij}=p_i\cdot p_j$ Lorentz products between particles' momenta. It can be initialised in several ways from the masses of the incoming and outgoing particles, the CoM energy and a pointer to \marty's numerical parameter structure:
\begin{lstlisting}
    Kinematics(std::vector<double> incoming_masses, std::vector<double> outgoing_masses, double sqrt_s, param_t* params)
    Kinematics(incoming_mass, std::vector<double> outgoing_masses, param_t* params)
\end{lstlisting}
where in the latter case, $\sqrt s$ is automatically set to the value of \texttt{incoming mass}. The \texttt{Kinematics} object then exposes a \texttt{update(std::vector<double> kin\_params)} function which automatically updates the values of the $s_{ij}$ products stored in the \texttt{param\_t} structure from a given set of kinematic parameters. The currently implemented kinematic distributions with their parameters are gathered in table~\ref{tab:kin}.

\begin{table}[ht]
    \centering
    \begin{tabular}{c|c|c}
        Kinematic type & Dimension & Parameters \\
        $1\to 2$ & 0 & - \\
        $1\to 3$ & 2 & $x=p_1\cdot p_2/m_1^2$, $y=p_1\cdot p_3/m_1^2$ \\
        $2\to 2$ & 1 & $\cos\theta$ \\
        $2\to 3$ & 4 & $t=(p_4+p_5)^2$, $\cos\theta$, $\cos\theta^*$, $\phi^*$
    \end{tabular}
    \caption{Natively implemented kinematics in \marty's numerical libraries and their set of kinematic parameters. $\theta$ is the angle between $p_2$ and $p_3$ in the collision's CoM, and $\theta^*,\phi^*$ are the polar and azimuthal angles of $p_4$ in the CoM of $p_3$.}
    \label{tab:kin}
\end{table}

The \texttt{Integrator} object takes care of the integration of a given squared amplitude over the corresponding phase space. It is initialised with the name of the function representing the squared amplitude, and an initialised \texttt{Kinematics} object encoding the kinematics of the process. 
\begin{lstlisting}
    Integrator(const std::string& f_M_squared, Kinematics& kinematics)
\end{lstlisting}
Once the \texttt{Integrator} is initialised, it exposes an \texttt{integrate()} method that computes the integral over the whole phase space using the adaptive Monte-Carlo algorithm \texttt{VEGAS} as implemented in the \texttt{GSL} library. It is not currently possible to apply phase-space cuts on the integration, as \marty\ is not intended to replace dedicated codes such as \amc~\cite{Alwall:2014hca}. After this step, one can call the method \texttt{get\_integral\_value()} to retrieve the value of the integral or \texttt{get\_integral\_error()} to retrieve the error estimation of \texttt{VEGAS} on the integral. The following sample code shows how to integrate the squared amplitude for the trident production process $\nu_\mu\gamma\to\nu_\mu\mu^-\mu^+$ in the EPA with a given CoM energy $\sqrt{s}$:

\begin{lstlisting}[language=c++,showstringspaces=false,upquote=false]
    param_t params;
    // Initialize all non-kinematic parameters in params
    Kinematics kin ({0, 0}, {0, m_mu, m_mu}, sqrt_s, &params);
    Integrator integrator ("M_nu_gamma__nu_mu_mu", kin);
    integrator.integrate()
    std::cout << "sigma(nu_mu gamma > nu_mu mu- mu+) @ sqrt_s = "
              << sqrt_s
              << " = " << integrator.get_integral_value() 
              << " +- " << integrator.get_integral_value() << "GeV^-2;
\end{lstlisting}

\paragraph{The case of on-shell $1\to3$ decays} For some values of $m_V$, the decays $\mu\to eee, \tau\to\ell\ell\ell, \mu\to e\nu\nu,\tau\to\ell\nu\nu$ can proceed through the on-shell production of a $V$ boson which decays into a pair of leptons afterwards. This shows up as poles in the integrand $f(x,y):=\Phi(x,y)|M(x,y)|^2$ (where $\Phi(x,y)$ is a phase-space factor) for all possible $V$ bosons and lepton arrangements. In practice, these poles are regulated by the small decay widths of the $V$ bosons and lead to denominators such as:
\begin{equation}
    f(x,y)\propto\frac{1}{\qty\big(m_V^2-m_\alpha^2-m_\beta^2+2p_\alpha\cdot p_\beta)^2+m_V^2\Gamma_V^2} \ ,
\end{equation}
which are sharply peaked around $p_\alpha\cdot p_\beta=s_{\alpha\beta}=(m_\alpha^2+m_\beta^2-m_V^2)/2$ (recall our integration variables are $x,y=p_1\cdot p_{2,3}/m_1^2$). Such sharp peaks require special attention, as none of the adaptive algorithms we tried could resolve them. Instead, we used the fact that the location of the peaks in $x$ and $y$  is known and that the peaks are Lorentzian in shape to decompose the integrand into: 
\begin{equation}
    f(x,y)=f_0(x,y)+L(x,\gamma_x)+L(y,\gamma_y) \ ,
\end{equation}
where $L$ is the Lorentz distribution: 
\begin{equation}
    L(t,\gamma)=\frac{L_0}{1+z^2}\qc z=\frac{t-t_0}{\gamma} \ ,
\end{equation}
where $t_0$ is the known location of the peak in $x$ or $y$, $L_0$ is the height of the peak which is estimated as: 
\begin{equation}
    L_0^x(y)=f(x_0,y)-\frac{f(x_0-n\gamma_x,y)+f(x_0+n\gamma_x,y)}{2}\qc n\sim10 \ ,
\end{equation}
and similarly for $L_0^y(x)$. The peak width $\gamma$ is then fitted over 20 points around the peak location. To speed up the fitting step, we use the mean decay width of the $V_p$ and $V_3$ bosons (which is also calculated automatically at tree-level by \marty) as an initial guess for $\gamma$ and to compute $L_0$. The remainder $f_0(x,y)$ is regular and can be readily integrated with the standard algorithms, and the contribution of the Lorentzian functions is known analytically and read:
\begin{equation}
    I_L=\int_{x_l}^{x_u} L(x,\gamma_x)dx=L_0^x\gamma_x\qty[\arctan(\frac{x_u-x_0}{\gamma_x})-\arctan(\frac{x_l-x_0}{\gamma_x})] \ .
\end{equation}
If $\gamma_x\ll |x_{u,l}-x_0|$ then $I_L\approx\pi L_0^x\gamma_x$.

\section{Dark matter interface: \darkpack and \darksusy}\label{chap:darkpack_darksusy_interface}

The calculation of the dark matter relic density in the model defined in \autoref{eqn:dm_lagrangian} has been performed by using both \darkpack and \darksusy.
In particular, the model whose implementation can be found in \autoref{chap:marty_impl} has been extended to add a Dirac fermion $SU(2)_f$ doublet, which is the dark matter candidate. 
In this case, \darkpack has been used to generate a numerical library which contains all the features to compute the dark matter relic density. 
However, we noticed that by using \darkpack's native method to obtain the plot of the coupling $g_f$ as a function of the mass of the neutral vector boson mediator $m_{V}$, the curve presented numerical instabilities. 
In order to solve this issue, the thermal relic density has been computed by using the Boltzmann-solver of \darksusy. 
In fact, \darkpack provides a friendly interface for being linked with other libraries, and \darksusy is structured modularly, so it easily accepts amplitudes from external libraries. 
Therefore, a plugin has been written for this purpose, which shall be released open-source in a separate work related to the upgrades of \darkpack. 

In particular, \darksusy requires as input the quantity $W_{\rm eff}(\sqrt{s})$ -- defined e.g.~in \cite{Gondolo:1998ah} -- whose function is called directly inside the Boltzmann-solver function, to compute the average annihilation cross-section $\langle \sigma v \rangle$ also defined in \cite{Gondolo:1998ah}: an essential ingredient for the resolution of the Boltzmann equation. 
\darksusy offers several algorithms to solve the Boltzmann equation in the freeze-out scenario.
The plot where $m_\chi / m_{V} = 1/3$ has been generated by using the algorithm number 20, i.e.~the recommended algorithm in \darksusy 6. This algorithm requires $W_{\rm eff}(\sqrt{s})$ a limited amount of times, since it tabulates its values. 
Hence, in this context, the plugin has been used to provide $W_{\rm eff}(\sqrt{s})$ by performing its full calculation in \darkpack.
However, the results with algorithm 20 when $m_\chi / m_{V} \geq 0.4 $ were not satisfactory, so the algorithm number 99 has been used, with a resulting much longer time, since it does not use any kind of tabulation for $W_{\rm eff}(\sqrt{s})$.
To circumvent the time constraint, an \textit{ad-hoc} function has been built in the plugin, which uses \darkpack's internal lookup table of $W_{\rm eff}(\sqrt{s})$ to provide very quickly the input for the algorithm 99 of the Boltzmann-solver. 
The results obtained as described have been used to use a minimisation method to find the points in the parameter space that reproduced the correct relic density.

The parameter space has been chosen to be 2-dimensional, as described in Sec.~\ref{sec:dm_limits}. Thus, the free parameters we considered are  $g_f$ and $m_V$. By choosing $m_V$ as the independent variable, the value of $g_f$ such that the model reproduces the correct relic density has been determined by using an adaptive bisection-secant method, which required approximately $20$ iterations per point, and a time between 20 and 40 minutes, depending on the chosen algorithm.

\section{Further numerical technical details}

\subsection{Nucleus electromagnetic form-factor }
The nucleus electromagnetic form-factor $F(q^2)$ used in Sec.~\ref{sec:trident} is related to the charge density of the nucleus by a Fourier transform:
\begin{equation}
    F(q)=\int_0^\infty\,r^2\rho(r)j_0(qr)\,\dd r \ ,
\end{equation}
where $j_0(x)$ is the first spherical Bessel function. There are several ways to parameterise the charge density of the nucleus. In our calculation, we used the Fourier-Bessel expansion (FBE):
\begin{equation}
    \rho(r)=\rho_0\sum_{k=1}^n a_kj_0\qty(\frac{k\pi r}{R}) \ ,
\end{equation}
for $r\leq R$ and $\rho(r>R)=0$. The coefficients $a_k$ and the radius $R$ are measured experimentally for iron which is the target of both CHARM and NuTeV.
The normalisation constant $\rho_0$ is fixed by normalising the total charge of the nucleus to 1. 
Taking kinematic limits into account to fix the integration bounds, the nuclear cross-section reads:
\begin{equation}
    \sigma(\nu N\to\nu\ell^+\ell^-)=\frac{2Z^2e^2}{\pi^2N^2}\int_{x_0}^\infty\frac{\dd x}{x}\int_{4m_\ell^2}^{s_{\rm max}}\frac{\dd s}{s}F^2(x)\sigma(\nu\gamma\to\nu\ell^+\ell^-) \ ,
\end{equation}
where
\begin{equation}
    x=\frac{\sqrt{q^2}R}{\pi}\ , \quad x_0=\frac{2Rm_\ell^2}{\pi E_\nu}\ , \quad s_{\rm max}=\frac{2\pi E_\nu x}{R} \ , 
\end{equation}
and $E_\nu$ is the energy of the incoming neutrino beam. In practice, due to the fast decay of $F(x)$, the $x$-integral can be truncated at $x_{\rm max}\approx 4$. As $F(x)$ varies much faster than the integrated EPA cross-section, in our numerical calculations the $s$-integral is computed precisely at 4 chosen points and then linearly interpolated at each $x$ value in the remaining integral. 

\subsection{Experimental efficiency in the long-lifetime limit}\label{chap:longlifetime}
In the small $m_V$ and small $g_f$ regime, the lifetime of the $V$ bosons becomes large enough for them not to decay promptly after being produced on-shell, leading to a decreasing efficiency for experimental searches requiring visible final states. For our limit estimation in the on-shell $V$ production regime, we applied an efficiency factor $\ev{\epsilon}(m_V)$ to the raw value of the lepton partial widths to take into account the loss in sensitivity due to displaced vertices of the form:
\begin{equation}
    \ev{\epsilon}=\ev{1-\exp(-\frac{L_{\rm prompt}}{L_V})}\ , \quad L_V=\frac{\beta\gamma}{\Gamma_V} \ ,
\end{equation}
where $L_{\rm prompt}$ is the cutoff length for a vertex to be considered prompt, and the boost factor $\beta\gamma$ is calculated within a toy Monte-Carlo simulation for each experiment which is used to find the averaged efficiency factor $  \ev{\epsilon}$.
\paragraph{SINDRUM} The $\mu\to eee$ limit comes from the SINDRUM experiment, a fixed-target muon-capture experiment with the muons thus decaying at rest. The boost factor of the produced $V$ boson is therefore: 
\begin{equation}
    \gamma=\frac{m_V^2+m_\mu^2}{2m_\mu m_V} \ .
\end{equation}
For our simulation we took a prompt decay length cutoff at $L_{\rm prompt}^{\rm SINDRUM}=\SI{1}{\milli\meter}$ (corresponding to interaction vertices that would be reconstructed outside of their target~\cite{SINDRUM:1987nra}).
\paragraph{BELLE} The case of $\tau\to\ell\ell\ell$ decays is more involved as it was measured by the BELLE detector at the KEKB collider, which operates an asymmetric electron beam to produce a pair of $\tau$ leptons which can then decay to $\ell V$. To be perfectly precise one should simulate the whole chain of boosts arising from the beam asymmetry, the kinematic distribution of the $\tau$ leptons in the rest frame of the collision, and the kinematic distribution of the $V$ bosons in the rest frame of the $\tau$ lepton. However, to percent-level precision one can consider that the decaying $\tau$ leptons are produced with an average boost of $\gamma_\tau=3$ in the lab frame, and only consider the boost of the $V$ boson from the rest frame of the $\tau$ (henceforth denoted by a $^*$) to the lab frame. The kinematic distribution of the boost factor in the lab frame then reads:
\begin{equation}
    \gamma=\frac{\gamma_\tau}{m_V}\qty(E_V^*+\beta_\tau p_V^*\cos\theta^*)\qc E_V^*=\frac{m_V^2+m_\tau^2}{2m_\tau}\qc p_V^*=\sqrt{{E_V^*}^2-m_V^2} \ .
\end{equation}
And we performed a toy Monte-Carlo simulation over $\cos\theta^*$ to compute the average efficiency from the expression of $\gamma$, and a prompt decay length cutoff at $L_{\rm prompt}^{\rm BELLE}=\SI{1}{\centi\meter}$ (as a rough estimate of the distance of the closest approach required in~\cite{Hayasaka:2010np}).

\end{appendices}

\bibliographystyle{utphys}
\bibliography{biblio.bib}

\end{document}